\documentclass[aps,pra,showpacs,twoside,twocolumn,10pt]{revtex4-2}
\usepackage[colorlinks=true, citecolor=red, urlcolor=blue ]{hyperref}
\usepackage{epsfig,newlfont,amssymb,amsfonts,amsmath,bm,subfigure,palatino,mathtools,amsthm,braket,soul,enumitem,color,graphics,graphicx,times,physics,bbold}
\usepackage[normalem]{ulem}
\usepackage{xcolor}
\usepackage{physics}
\usepackage{dsfont}
\usepackage{mathrsfs}
\usepackage{verbatim}
\usepackage{amsthm,amssymb}
\usepackage{comment}
\usepackage{bigints}

\begin{document}

\AtBeginDocument{
  \renewcommand{\Re}{\mathfrak{R}}
  \renewcommand{\Im}{\mathfrak{I}}
}

\title{ Non-Markovian and Thermodynamic Signatures in the Classicality Assessment\\ via Kolmogorov Consistency
}


\author{Arghya Maity$^{1}$, Ahana Ghoshal$^{2}$, Kelvin Onggadinata$^{1}$, Teck Seng Koh$^{1}$}
\affiliation{$^{1}$  School of Physical and Mathematical Sciences, Nanyang Technological University, 21 Nanyang Link, Singapore 637371, Singapore\\
$^{2}$ Naturwissenschaftlich-Technische Fakult\"{a}t, Universit\"{a}t Siegen, Walter-Flex-Stra\ss e 3, 57068 Siegen, Germany}

\begin{abstract} 
The Kolmogorov consistency condition (KCC) defines the statistical boundary between classical and quantum dynamics. Its violation signifies the breakdown of a classical Markov description of temporal correlations. In this work, we establish a direct analytical connection between KCC violation and non-Markovianity in open quantum dynamics, revealing how memory effects manifest as departures from classical probabilistic consistency. 
Within a generic two-level open quantum system framework, we establish quantitative connections between the magnitude of KCC violation and key information-theoretic and thermodynamic quantities, such as mutual information, the Fano factor, heat exchange, and entropy production rate, thereby enabling a thermodynamic interpretation of temporal quantum correlations.
Furthermore, we uncover formal correspondences between KCC violation, the
Leggett--Garg inequality, and the negativity of the Kirkwood--Dirac
quasi-distribution, identifying them as complementary witnesses of temporal
quantum non-classicality. Our results thus provide a unified framework linking
information-theoretic, thermodynamic, and temporal indicators of quantumness in
open quantum systems.
\end{abstract}

\maketitle

\section{Introduction}
Understanding the boundary between classical and quantum dynamics remains a central challenge in physics~\cite{Engel_Nature_2007,Huelga_CP_2013,Zambrini_PRL_2014,Lostaglio_RPP_2019,Correa_PRE_2019,Wang_NRC_2019}. Distinguishing genuinely quantum behaviour from classical stochastic evolution is crucial for revealing the fundamental nature of dynamical processes and for advancing quantum technologies, where coherence, noise, and memory effects determine operational limits.

Several theoretical frameworks have been developed to explain how classicality emerges from quantum mechanics~\cite{Zurek_RMP_2003,Joos_2013,Schlosshauer_2007,Schlosshauer_PR_2019,Kent_JSP_1996,Griffiths_2019_SEP,Gemmer_PRE_2016,Gemmer_PRE_2014,Smirne_QST_2019,Strasberg_PRA_2019,Kavan_PRXQuantum_2021,Huelga_PRX_2020,Strasberg_2022,Kavan_Quantum_2020,Strasberg_PRA_2023,Gambini_GRG_2007,sep-qm-collapse,Strasberg_SP_2023,Leggett_JP_2002,Daniel_PRD_2021,Zurek_Nature_2009,Korbicz_Quantum_2021,Zurek_Entropy_2022}. 
Among these, the open-quantum-system perspective~\cite{Zurek_RMP_2003,Joos_2013,Schlosshauer_2007,Schlosshauer_PR_2019,Strasberg_SP_2023} attributes classical behaviour to decoherence, the rapid suppression of coherences in a preferred (pointer) basis due to environmental coupling~\cite{Zurek_RMP_2003,Joos_2013,Schlosshauer_2007,Schlosshauer_PR_2019,Strasberg_SP_2023,Petruccione_book,Vega_RPM_2017}. After a short decoherence time, the system’s reduced state becomes effectively diagonal in this basis, enabling classical-like statistics. However, decoherence alone does not ensure that multi-time probabilities satisfy classical consistency relations; even when coherences vanish, measurement backaction and residual system–environment correlations can lead to violations of classical stochasticity~\cite{Strasberg_SP_2023}. 
This observation holds across different theoretical approaches to the emergence of classicality. In particular, since decoherence in open quantum systems is not sufficient to ensure Kolmogorov consistency, it likewise cannot account for classicality within the decoherent-histories framework~\cite{Kent_JSP_1996,Griffiths_2019_SEP}, which characterizes classical behaviour through the consistency of entire sequences of quantum events. Interestingly, the additional assumption of Markovianity is sufficient to establish that open-system decoherence leads to decoherent histories~\cite{Strasberg_SP_2023}.

A more stringent requirement for classical behaviour is expressed by the Kolmogorov consistency condition (KCC)~\cite{Kolmogorov_1950,Petruccione_book}, which demands that multi-time probability distributions remain mutually consistent when marginalized over intermediate variables. Any violation of KCC indicates the absence of an underlying classical stochastic process and thus signals non-classical evolution. In quantum systems, such violations naturally arise from coherence, memory effects, or environmental feedback, making KCC a powerful operational tool to identify non-classicality. When the dynamics become non-Markovian, memory effects and information backflow can lead to violations of the Kolmogorov consistency condition, revealing a breakdown of classical probabilistic structure.

Within the Markovian regime, where dynamics are memoryless and independent of past interventions, KCC is automatically fulfilled. Indeed, for Markovian quantum processes, a one-to-one correspondence between Kolmogorov consistency and dynamical coherence has been established~\cite{Smirne_QST_2019}. This understanding has been partially extended to the non-Markovian domain—where information backflow and memory effects occur~\cite{Huelga_PRX_2020}. However, since non-Markovianity can manifest in various forms and be characterized through multiple approaches, it remains valuable to explore its different facets to uncover new and physically significant insights. In realistic scenarios, most quantum systems evolve non-Markovianly, making it essential to determine under what conditions consistent, effectively classical statistics can still arise.
Interestingly, recent studies have shown that classical-like behaviour can also emerge without explicit decoherence. Isolated many-body systems, even when highly coherent, may exhibit effectively classical statistics due to non-integrability and chaos, which suppress observable quantum effects for coarse and slow observables~\cite{Gemmer_PRE_2014,Strasberg_PRA_2023,Gemmer_PRE_2016}. These results suggest that both decoherence and dynamical complexity contribute to the emergence of classicality, while non-Markovian memory introduces new pathways for its breakdown.

In open quantum systems, non-classical features such as information backflow, temporal correlations, and memory-induced coherence revivals play a pivotal role. Understanding these effects provides insight into the interplay between information and thermodynamics—linking KCC violations to quantities such as entropy production, heat flow, and irreversibility. Establishing quantitative connections between the degree of KCC violation and these thermodynamic observables can reveal how information-theoretic and energetic aspects of quantum evolution are intertwined.

In this work, we employ an exact master-equation approach to analytically evaluate the violation of the Kolmogorov consistency (KCC) condition, revealing how classicality breaks down in open quantum systems and under what conditions this breakdown is amplified. The analytical results demonstrate how the degree of KCC violation depends on the interplay between positive and negative decay rates, thereby linking three fundamental aspects of open quantum dynamics --- Hamiltonian evolution, dissipation, and memory effects. We show that the emergence of KCC violation serves as an operational indicator of non-classical stochastic behavior induced by non-Markovian evolution. 
Furthermore, we establish explicit connections between KCC violation and standard non-Markovianity measures such as the RHP (CP-divisibility)~\cite{Rivas_PRL_2010,Rivas_RPP_2014} and BLP (information backflow)~\cite{BLP_PRL_2009,BLP_PRA_2010, onggadinata2025efficient} quantifiers. In particular, the KCC violation exhibits exponential suppression governed by positive decay rates, while negative-rate intervals —- responsible for both RHP non-divisibility and BLP revivals —- act as amplification channels.

We further investigate the correspondence between KCC violation with information-theoretic and thermodynamic quantities, including mutual information, heat flow, entropy production. To clarify its physical significance, we explore whether this temporal non-classicality is linked to thermodynamic irreversibility. In certain regimes, we show that the exchanged heat and entropy-production rate can act as thermodynamic witnesses of KCC violation. Overall, our analysis provides a unified framework connecting information dynamics, thermodynamics, and non-Markovian memory effects in the characterization of quantum temporal correlations.


Another fundamental viewpoint on temporal non-classicality is provided by the
Leggett--Garg inequalities (LGIs)~\cite{Leggett_Garg_PRL_1985,Leggett_RPP_2008},
which test macrorealism—the assumption that a system possesses definite
properties at all times, independent of measurement. A closely related notion is
embedded in the Kolmogorov consistency condition (KCC), which requires that
joint probabilities across different times remain consistent with their
marginals. Since both LGIs and KCC are derived from the same classical
assumptions of temporal realism and non-invasiveness, any dynamics obeying KCC
automatically satisfies LGIs~\cite{Asano_PS_2014,Emary_RPP_2014}. Therefore, a
violation of either condition signals a departure from classical stochastic
descriptions and reveals genuine quantum temporal correlations. In a broader
context, the Kirkwood--Dirac (KD) quasi-distribution provides a complementary
perspective on sequential quantum measurements, where its negativity directly
captures non-classical interference effects~\cite{Kirkwood_PR_1933,Dirac_RPM_1945}.
In this work, we establish explicit relationships among the degree of KCC
violation, KD-negativity, and LGI violation in a dissipative qubit model. Our
results demonstrate that these distinct manifestations of temporal
quantumness share a common physical origin: coherence revivals driven by
non-Markovian memory effects associated with temporarily negative decay rates.

The remainder of the paper is organized as follows. In Sec.~\ref{KCV_Model}, we introduce the mathematical formulation of Kolmogorov consistency and its violation. We then present the exact master equation model that we use to study non-Markovian dynamics, in Sec. \ref{Master_eq}, followed by two specific models that allow us to obtain both analytical and numerical results in a clear and accessible manner in Sec. \ref{model} and \ref{cases}. In Secs.~\ref{KC_RHP} and \ref{KC_BLP}, we relate the violation of Kolmogorov consistency to two measures of non-Markovianity, namely the RHP and BLP measures. We further examine its correspondence with information-theoretic and thermodynamic quantities, including mutual information in Sec.~\ref{KC_MI}, as well as heat flow and entropy production in Sec.~\ref{KC_Q_sigma}. In Sec.~\ref{KC_Fano}, we connect Kolmogorov consistency violation to the Fano factor. Finally, in Secs.~\ref{KC_KD} and \ref{KC_LGI}, we discuss its formal links with the Leggett–Garg inequality and with the negativity of the Kirkwood–Dirac quasi-distribution, before concluding the work in Sec.~\ref{Con}.


\section{Kolmogorov Consistency and its Violation}
\label{KCV_Model}

The Kolmogorov extension theorem guaranties the existence of a well-defined classical stochastic process when the family of multi-time joint probability distributions satisfies the Kolmogorov consistency condition. For a sequence of $n$-measurement outcomes \(x_1, x_2, \ldots, x_n\) at times \(t^\prime_1 < t^\prime_2 < \ldots < t^\prime_n\), this condition requires that marginalizing over any subset of variables yields the correct joint probabilities for the remaining outcomes. Specifically, for any $k \leq n$,
\begin{align}
\sum_{x_k} P(x_n, \ldots, x_{k+1}, x_k, x_{k-1}, \ldots, x_1) \nonumber\\ 
= P(x_n, \ldots, x_{k+1}, x_{k-1}, \ldots, x_1).
\label{eq:KC_general}
\end{align}

Any deviation from the Kolmogorov consistency condition indicates a KCC violation. For simplicity, one can consider the two-time case $(t^\prime_1,t^\prime_2)$, where the violation is quantified as
\begin{equation}
\text{viol}_{x_2}(t^\prime_1,t^\prime_2) =
\left| \sum_{x_{1}} P(x_{2},x_{1}) - P(x_{2}) \right|.
\label{Eq:KC_viol_gen}
\end{equation}
Here, $P(x_2,x_1)$ denotes the joint probability of obtaining the measurement outcome $x_1$ at time $t^\prime_1$ and $x_2$ at time $t^\prime_2$, while $P(x_2)$ represents the single-time probability of getting the outcome $x_2$ at $t^\prime_2$ when no prior measurement is performed at $t^\prime_1$. 
A nonzero value, $\text{viol}_{x_2}(t^\prime_1,t^\prime_2) > 0$, signals the breakdown of a classical stochastic description.
For two-time probabilities, there are in principle two possible marginal consistency
conditions. However, marginalization over the later-time outcome $x_2$ trivially
reproduces the statistics at $t'_1$, while marginalization over the earlier outcome
$x_1$ provides a nontrivial test of measurement-induced disturbance. We therefore
restrict our attention to the latter, as expressed in Eq.~\ref{Eq:KC_viol_gen}.
Classical evolution implicitly assumes a stable pointer basis, in which measurements do not disturb the dynamics and joint probabilities remain well defined. Decoherence naturally selects such a basis by suppressing quantum coherences, thereby ensuring that the Kolmogorov consistency condition is satisfied.

In contrast, in non-Markovian dynamics, information can flow back from the environment to the system, leading to deviations from classical stochastic behavior. Such deviations can be captured by KCC violations. In the two-time case, this reads
$
\sum_{x_{1}} P(x_{2},x_{1}) \ne P(x_{2}) ,
$
highlighting a breakdown of classical stochasticity. In the present work, we focus on non-Markovian dynamics arising when a time-local master equation develops negative decay rates and aim to establish a direct connection between KCC violation and non-Markovianity.



\section{Time-Local Master Equations with Negative Decay Rates}
\label{Master_eq}
We consider a quantum system initialized in the state $\rho(0)$, evolving under the system Hamiltonian $H_{\text{sys}}$ and coupled to an external reservoir. At the beginning of the evolution, the reservoir is assumed to be in a thermal equilibrium state given by $\rho_{R}=\frac{e^{-\beta^\prime H_R}}{\text{Tr}\big(e^{-\beta^\prime H_{R}}\big)}$, where $H_R$ is the local Hamiltonian of the reservoir and $\beta^\prime=1/k_B T$ is the inverse temperature of the reservoir with $k_B$ being the Boltzmann constant. Furthermore, at the initial time, the system and the reservoir are assumed to be uncorrelated, such that the joint system–bath state can be expressed as a direct product, $\rho_{sR}(0)=\rho(0)\otimes \rho_R$. With these initial assumptions, the dynamics of the system can be given by the exact time-local master equation
\begin{equation}
\label{Eq:dyn_equ}
\frac{d\rho(t^\prime)}{dt^\prime} = -\frac{i}{\hbar}\big[H_{\text{sys}}, \rho(t^\prime)\big] + \mathcal{D}[\rho(t^\prime)],
\end{equation}
where the first term describes the unitary evolution of the system, and $\mathcal{D}[\rho(t^\prime)]$ denotes the dissipator, incorporating both dissipation and fluctuation effects arising from the system–reservoir interaction. The dissipator takes the form

\begin{align}
\label{Eq:Master_Equ}
\mathcal{D}[\rho(t^\prime)] = \sum_{\varepsilon\ge 0}
\Gamma^\prime_{\varepsilon}(t^\prime)\,L_{A(\varepsilon)}[\rho(t^\prime)]+ \widetilde{\gamma}^\prime_{\varepsilon}(t^\prime)\,
L_{A^{\dagger}(\varepsilon)}[\rho(t^\prime)],
\end{align}
with
\begin{equation}
L_{A(\varepsilon)}[\rho(t^\prime)]
= A(\varepsilon)\rho(t^\prime)A^{\dagger}(\varepsilon)
-\tfrac{1}{2}\{A^{\dagger}(\varepsilon)A(\varepsilon),\rho(t^\prime)\}, 
\end{equation}
where $A(\varepsilon)$ are the jump operators associated with transitions between energy eigenstates differing by energy $\hbar \varepsilon$. $\Gamma^{\prime}_{\varepsilon}(t^{\prime})$ and $\widetilde{\gamma}^{\prime}_{\varepsilon}(t^{\prime})$ are time-dependent decay rates determined by the spectral properties of the reservoir and its correlation functions. When the rates $\Gamma_{\varepsilon}(t)$ and $\widetilde{\gamma}_{\varepsilon}(t)$ remain positive for all times, the dynamics are Markovian and completely positive. However, when these rates temporarily become negative, the dynamics become non-Markovian, signaling the presence of information backflow from the environment to the system.

Under weak system-reservoir coupling and retaining terms up to second order in the system-reservoir coupling strength, the decay rates take the approximate forms~\cite{Nori_PRL_2012,Xiong_PRA_2010}
\begin{align}
\gamma^\prime_{\varepsilon}(t^{\prime}) &\approx
\int_{0}^{t^{\prime}} ds \int \tfrac{d\omega'}{2\pi}\,
\mathcal{J}(\omega')\cos[(\omega'-\varepsilon)(t^{\prime}-s)], \nonumber \\
\widetilde{\gamma}_{\varepsilon}
(t^{\prime}) &\approx
2\int_{0}^{t^{\prime}} ds \int \tfrac{d\omega'}{2\pi}\,
\mathcal{J}(\omega')
n(\omega')\cos[(\omega'-\varepsilon)(t^{\prime}-s)],
\label{gamma}
\end{align}
where $\Gamma^{\prime}_{\varepsilon}(t^{\prime}) = 2\gamma_{\varepsilon}(t^{\prime}) \mp \widetilde{\gamma}_{\varepsilon}(t^{\prime})$ and $n(\omega')=[e^{\beta^{\prime}\hbar\omega'}\mp 1]^{-1}$ with the upper (lower) sign corresponds to bosonic (fermionic) reservoirs.
\section{The model}
\label{model}
We consider a single-qubit system described by the Hamiltonian
\begin{equation}
    H_{\text{sys}}=\frac{\hbar}{2}\omega_0 \sigma_z,
\end{equation}
where $\omega_0$ is the qubit transition frequency and $\sigma_z$ is the Pauli-$z$ matrix.
The system is coupled to a bosonic reservoir described by the Hamiltonian
\begin{equation}
    H_R = 
\hbar \omega^{\prime}_c\int^{\infty}_{0} d\omega^{\prime} a^{\dagger}_{\omega^{\prime}} a_{\omega^{\prime}},
\end{equation}
where $a^{\dagger}_{\omega^{\prime}}$ and $a_{\omega^{\prime}}$ denote, respectively, the creation and annihilation operators of the reservoir mode with frequency $\omega^\prime$. These operators satisfy the canonical commutation relation $[a_{\omega^{\prime}},a_{\omega^{\prime\prime}}^{\dagger}]=\delta(\omega^{\prime}-\omega^{\prime\prime})$. $\omega_c$ is the cutoff frequency that determines the memory timescale of the reservoir. The interaction between the system and the reservoir is modeled as
\begin{equation}
    H_I=\hbar \sqrt{\omega^{\prime}_c} \int_{0}^{\infty} d\omega^{\prime} \chi(\omega^{\prime}) \Big( \sigma_{+} a_{\omega^{\prime}} + \sigma_{-} a^{\dagger}_{\omega^{\prime}}  \Big),
\end{equation}
where $\sigma_{\pm} = (\sigma_{x} \pm i \sigma_{y}) / 2$ are the qubit raising and lowering operators. The dimensionless function
$\chi(\omega^{\prime})$ characterizes the frequency-dependent coupling strength between the qubit and the reservoir. It is related to the spectral density function of the reservoir, $\mathcal{J}(\omega^{\prime})$, through $2\pi\omega_c|\chi(\omega^{\prime})|^2=\mathcal{J}(\omega^{\prime})$. To describe the environmental structure, we adopt a super-Ohmic spectral density of the form
$\mathcal{J}(\omega) = \alpha\,\omega^{s}e^{-\omega^\prime/\omega^{\prime}_c}$, with $s> 1$.
For this scenario, the dynamical equation of the system reduces to
\begin{equation}
    \frac{d\rho(t^{\prime})}{dt^{\prime}} = -\frac{i}{\hbar}\big[H_{\text{sys}}, \rho(t^{\prime})\big] + \Gamma^{\prime}_{\varepsilon}(t^{\prime})\,L_{\sigma_-}[\rho(t^{\prime})]+ \widetilde{\gamma}^{\prime}_{\varepsilon}(t^{\prime})\,
L_{\sigma_+}[\rho(t^{\prime})].
    \label{Eq:Master_Eq1}
\end{equation}
To simplify the analysis, we introduce dimensionless variables by measuring all
frequencies in units of the qubit frequency $\omega_0$ and time in units of
$\omega_0^{-1}$. We define
\begin{align}
     t = \omega_0 t^{\prime}, &\qquad
    \Gamma( t)=\frac{\Gamma^{\prime}(t^{\prime})}{\omega_0},
    \qquad
    \widetilde{\gamma}(t)=\frac{\widetilde{\gamma}^{\prime}(t^{\prime})}{\omega_0}, \nonumber \\
    &\omega_c=\frac{\omega^{\prime}_c}{\omega_0},\qquad
    \beta=\beta^{\prime}\hbar\omega_0 \nonumber.
\end{align}

\section{Kolmogorov--consistency condition for a single-qubit system}
\label{cases}
We consider the system as a single qubit whose state at time $t$ is represented by the density matrix
\begin{align}
\rho(t)=\begin{pmatrix} a(t) & c(t) \\ c^*(t) & 1-a(t) \end{pmatrix}.
\label{eq:state1}
\end{align}
To examine the validity of the Kolmogorov consistency condition, we must first specify the measurement basis in which the measurements are performed.
We consider a general orthonormal measurement basis defined as
\begin{align}
|u_1\rangle=&\cos\frac{\theta}{2}\,|0\rangle+e^{i\phi}\sin\frac{\theta}{2}\,|1\rangle,
\nonumber \\
|u_2\rangle=&-e^{-i\phi}\sin\frac{\theta}{2}\,|0\rangle+\cos\frac{\theta}{2}\,|1\rangle,
\end{align}
where $0\le \theta \le \pi$ and $0 \le \phi < 2\pi$. We now need to determine two kinds of probabilities: the single-time probabilities and the conditional probabilities. The single-time probabilities are obtained by letting the system evolve from its initial state up to the measurement time and then performing a projective measurement. The conditional probabilities are obtained by first measuring the system at an intermediate time, allowing the post-measurement state to evolve further, and then performing a second measurement.

To check the validity of Kolmogorov consistency condition, we evaluate the quantity $\text{viol}(t_1,t_2)$ for the outcome $|u_1\rangle$. In this case, it is defined as
\begin{equation}
\label{Eq:viol}
    \text{viol}_{u_1}(t_1,t_2)=\big |p_{u_1}(t_2)- \Pi_{u_1}(t_2) \big |. 
\end{equation}
Here, $p_{u_1}(t_2)$ denotes the probability of obtaining the outcome $|u_1\rangle$ when a single measurement is performed at time $t_2$ without any prior measurement and $\Pi_{u_1}(t_2)$ represents the classical mixture of conditional probabilities, defined as
\begin{equation}
    \Pi_{u_1}(t_2) := p_{u_1}(t_1)\,p_{u_1}(t_2\!\mid\!u_1) + p_{u_2}(t_1)\,p_{u_1}(t_2\!\mid\!u_2),
\end{equation}
where $p_{u_i}(t_1)$ is the probability of obtaining the outcome $|u_i\rangle$ at time $t_1$, and $p_{u_1}(t_2\!\mid\!u_i)$ is the conditional probability of obtaining $|u_1\rangle$ at time $t_2$ given that the outcome $|u_i\rangle$ was obtained at $t_1$. The single-time probabilities $p_{u_i}(t_i)$ are given by
\begin{eqnarray*}
    &&p_{u_1}(t_i)=\tfrac{1}{2}+\big(a(t_i)-\tfrac{1}{2}\big )\cos \theta
+ \sin \theta\,\Re\big[e^{i\phi}c(t_i)\big], \nonumber\\
&&p_{u_2}(t_i)=1-p_{u_1}(t_i),
\end{eqnarray*}
and the conditional probabilities $p_{u_1}(t_2|u_i)$ are
\begin{equation*}
p_{u_1}(t_2\!\mid\!u_i)
= \tfrac{1}{2} + \big(a_{u_i}(t_2)-\tfrac{1}{2}\big)\cos\theta
+\sin\theta\, \Re\!\big[e^{i\phi}c_{u_i}(t_2)\big].
\end{equation*}
Here,  $a_{u_i}(t_2)$ and $c_{u_i}(t_2)$ represent the population and coherence elements of the post-measurement state that has evolved from $|u_i\rangle$ at time $t_1$ to $t_2$. Substituting these expressions of the probabilitites in Eq.~(\ref{Eq:viol}), we get
$$
\mathrm{viol}_{u_1}(t_1,t_2)
= \big|\,\mathcal{P}(t_1,t_2)\;+\;\mathcal{C}(t_1,t_2)\,\big|,
$$
where
$$
\begin{aligned}
\mathcal{P}(t_1,t_2)
&= \cos\theta\Big\{\big(a(t_2)-\tfrac{1}{2}\big)
-\sum_{i=1}^2 p_{u_i}(t_1)\big(a_{u_i}(t_2)-\tfrac{1}{2}\big)\Big\},\\[6pt]
\mathcal{C}(t_1,t_2)
&= \sin\theta\Big\{\Re\big[e^{i\phi}c(t_2)\big ] - \sum_{i=1}^2 p_{u_i}(t_1)\Re\!\big[e^{i\phi}c_{u_i}(t_2)\big]\Big\}.
\end{aligned}
$$
The presence of a nonzero $\text{viol}(t_1, t_2)$ thus signifies a breakdown of Kolmogorov consistency, reflecting the influence of non-classicality in the dynamics. See App.~\ref{app:1} for the detailed calculation of the single-time and conditional probabilities, as well as the Kolmogorov-consistency violation term $\text{viol}(t_1,t_2)$. To illustrate the violation more explicitly, we consider two special but ubiquitous cases that allow a detailed analysis of the nature of the KCC violation and its connection to relevant thermodynamic properties. 
\\
\\
\textbf{Case I : }
We consider the system to be initially prepared in the state $\rho(0)=|\psi(0)\rangle\langle \psi(0)|$, with
\[
|\psi(0)\rangle = |+\rangle = \tfrac{1}{\sqrt{2}}\big(|0\rangle + |1\rangle\big),
\]
and the measurements are performed in the computational basis $\{|+\rangle, |-\rangle\}$. This scenario corresponds to the parameters $a(0)=c(0)=\frac{1}{2}$, $\theta=\frac{\pi}{2}$, and $\phi=0$.
The system–bath interaction naturally selects the computational basis $\{|0\rangle, |1\rangle \}$ as the pointer basis. In contrast, performing measurements in the $\{|+\rangle, |-\rangle\}$ basis deliberately departs from this preferred basis, thereby enabling the manifestation of KCC violation.
For this case, the Kolmogorov-consistency violation for outcome $|+\rangle$ is
\begin{align}
\mathrm{viol}_+(t_1,t_2) 
= \tfrac12\,\exp\!&\Big(-\tfrac12 G(0,t_2)\Big) \\
&\big|\sin(\omega_0 t_1)\sin(\omega_0(t_2-t_1))\big|\nonumber
\end{align}
with $G(t_0,t):=\int_{t_0}^{t}\lambda(s)\,ds$ and,
$\lambda(s):=\Gamma_{\varepsilon}(s)+\widetilde{\gamma}_{\varepsilon}(s)$.
Substituting the expression for the time-dependent decay rate $\lambda(s)$, we finally obtain
\begin{align}
\mathrm{viol}_{+}(t_1,t_2)
&= \tfrac12\exp\!\Big(-\tfrac12\int_0^{t_2}\Big[\Gamma_{\varepsilon}(s)+\widetilde{\gamma}_{\varepsilon}(s)\Big]ds\Big) \nonumber
\\
&\quad \quad \quad \big|\sin(\omega_0 t_1)\sin(\omega_0(t_2-t_1))\big|.
\label{Eq:viol2}
\end{align}
See App.~\ref{app:1} for the detailed calculation.
Since the setup is symmetric with respect to the measurement outcomes $|+\rangle$ and $|-\rangle$, the same result holds for both cases, i.e., $\text{viol}_+(t_1,t_2)=\text{viol}_-(t_1,t_2)$.  
For notational simplicity, we therefore drop the outcome subscript and
denote the violation as $\mathrm{viol}(t_1,t_2)$.

To reveal the effects of positive and negative decay-rate contributions, we decompose any real function $x(t)$ into its positive and negative parts, $x(t)=x^+(t)-x^-(t)$ with $x^\pm(t)\ge0$.  Applying this to the total decay rate $\lambda(t)=\Gamma_\varepsilon(t)+\widetilde{\gamma}_{\varepsilon}(t)$ with $\lambda^{\pm}(t)=\Gamma^{\pm}_{\varepsilon}(t)+\widetilde{\gamma}_{\varepsilon}^{\pm}(t)$, we define
\begin{align}
\label{Eq:lambda}
&\phantom{ami pothbhola ek}\lambda(t)=\lambda^+(t)-\lambda^-(t), \nonumber\\
&M(t_2):=\int_0^{t_2}\lambda^+(s)\,ds , 
~~~~~N(t_2):=\int_0^{t_2}\lambda^-(s)\,ds,
\end{align}
so that $\int_0^{t_2}\lambda(s)\,ds=M(t_2)-N(t_2)$. Substituting into Eq.~(\ref{Eq:viol2}) yields the factorized form
\begin{align}
    \label{Eq:viol_MN}
     \mathrm{viol}_+(t_1,t_2) 
 =  \tfrac12\,e^{-\tfrac12 M(t_2)}&\big|\sin(\omega_0 t_1)\sin(\omega_0(t_2-t_1))\big|\nonumber\\
&\phantom{ki name daki}\times e^{+\tfrac12 N(t_2)}.
\end{align}
Here, the factor involving the sine terms originate solely from the unitary dynamics governed by the system Hamiltonian and capture the coherent oscillations of measurement statistics due to the choice of a non-commuting measurement basis.  
The KCC violation vanishes when either $\sin(\omega_0t_1)=0$ or $\sin(\omega_0(t_2-t_1))=0$, i.e., when $t_1$ or the interval $\Delta=t_2-t_1$ equals an integer multiple of $\pi/\omega_0$.
\begin{figure*}[htb]
\includegraphics[width=1.0\textwidth]{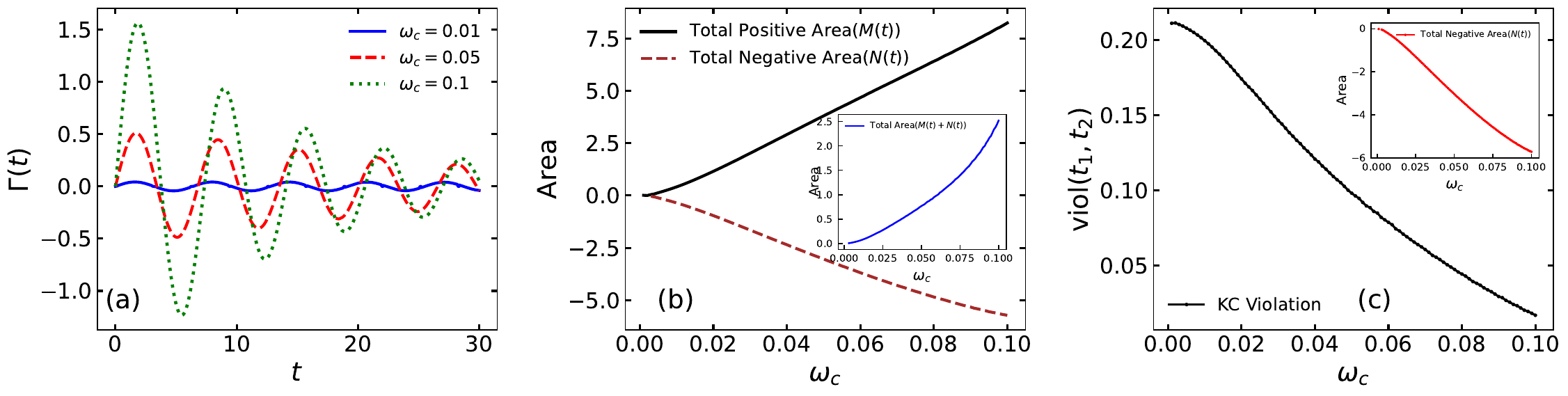}
\caption{Analytical results for Case I. The spectral density exponent is fixed at $s=1.5$, the system--bath coupling strength at $\alpha=0.5$, and the bath temperature at $T=300$, while the system Hamiltonian frequency is $\omega_0=1.0$.
To make every parameter dimensionless, $t'=\omega_0t$. The $\Gamma'(t)=\Gamma(t')/\omega_0$. 
(a) Time dependence of the decoherence rate $\Gamma(t)$ for different cutoff frequencies $\omega_c$. Varying $\omega_c$ significantly modifies the structure of $\Gamma(t)$, in particular the extent of its negative regions, thereby allowing control over the degree of non-Markovianity.
(b) Total positive area $M(t)$ and total negative area $N(t)$ of the decay rate, evaluated over a long time window ($t=30$), as functions of $\omega_c$. The positive contribution dominates for all cutoff frequencies, leading to an overall increase of the net area, as highlighted in the inset.
(c) Kolmogorov-consistency violation $\mathrm{viol}(t_1,t_2)$ as a function of $\omega_c$. The measurement is done at $t_1=15.0$ and the second at $t_2=30.0$. In agreement with the analytical expression Eq.~\eqref{Eq:viol_MN}, the violation decreases with increasing cutoff frequency, reflecting the reduction of the negative-rate contributions, as also illustrated in the inset. 
}
\label{Fig:Gamma_Area_KCV}
\end{figure*}
Otherwise, the violation becomes nonzero, and Eq.~(\ref{Eq:viol_MN}) reveals how the classical Kolmogorov consistency is disrupted, as well as the conditions under which this disruption can be amplified. The first exponential term
$e^{-\tfrac12 M(t_2)}$ accounts for the decay of coherence due to positive decay rates, while the last exponential term $e^{+\tfrac12 N(t_2)}$ quantifies the enhancement of the violation arising from the non-Markovian backflow of information associated with negative decay-rate contributions. 
Hence, the KC violation embodies a competition between two opposing mechanisms: (i) Markovian damping that promotes classical consistency and macrorealism, and (ii) non-Markovian backflow that restores coherence and revives non-classical statistics. 

An immediate consequence of the analytical form of the violation in Eq.~(\ref{Eq:viol_MN}) is the upper bound
\[
\mathrm{viol}_+(t_1,t_2)\le \frac{1}{2}\,e^{-\tfrac12\!\left[M(t_2)-N(t_2)\right]},
\]
which follows from $|\sin(\omega_0 t_1)\sin(\omega_0(t_2-t_1))|\le1$. 
It is worth noting that if $N(t_2) > M(t_2)$ the exponent in the bound becomes positive, so the right-hand side may formally exceed $1/2$. However, the overall violation remains limited by the sinusoidal prefactor, and very large $N(t_2)$ typically indicates regimes where the time-local description or complete positivity may break down.



To validate our analytical results, we perform numerical simulations and compare the trends with our theoretical predictions.

\begin{figure}[htb]
\includegraphics[width=0.4\textwidth]{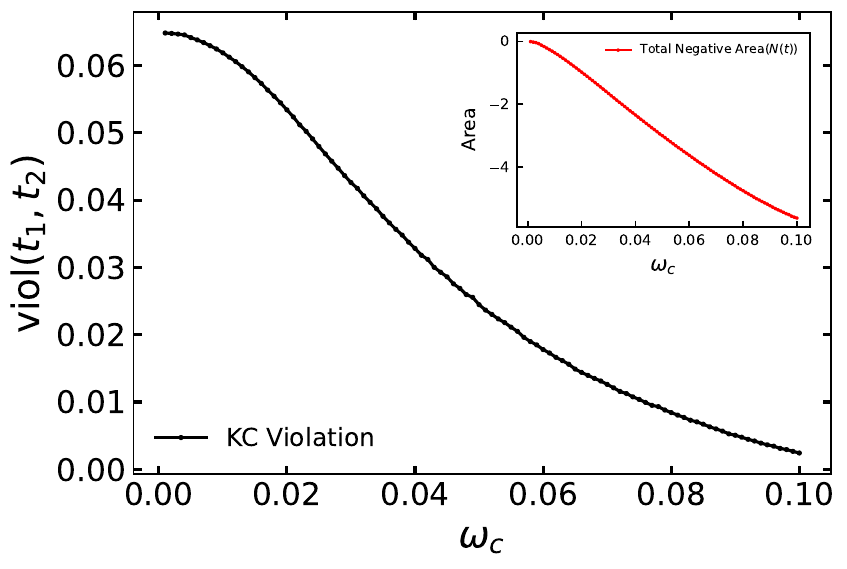}
\caption{Analytical results for Case II. All system and bath parameters are identical to those used in Fig.~\ref{Fig:Gamma_Area_KCV}(a), so the positive and negative areas of the decay rates coincide with those of Case I. The total evolution time and the measurement times are also chosen to be the same. The KC violation exhibits the same qualitative dependence on the positive and negative decay-rate areas, confirming that its behavior is governed by their interplay rather than by case-specific details.}
\label{Fig:model2_KCV}
\end{figure}

Figure~\ref{Fig:Gamma_Area_KCV}(a) illustrates the time evolution of the decoherence rate $\Gamma(t)$ for different values of the environmental cutoff frequency $\omega_c$ in the spectral density function. Correspondingly, in Fig.~\ref{Fig:Gamma_Area_KCV}(b), we show how the total positive area ($M$) and total negative area ($N$) of the decay rates vary with $\omega_c$. The inset further displays the total area ($M+N$), which increases with increasing $\omega_c$ due to the dominance of $M$ over $N$ for all $\omega_c$.
As a consequence, the temporal violation $\text{viol}(t_1,t_2)$, presented in Fig.~\ref{Fig:Gamma_Area_KCV}(c), decreases as $\omega_c$ increases, consistent with Eq.~\ref{Eq:viol_MN}. This relation also explains the trend seen in the inset of Fig.~\ref{Fig:Gamma_Area_KCV}(c): the violation diminishes as the negative contribution $N$ decreases.
\\
\\
\textbf{The Case II: } For further numerical investigation, we now consider an alternative model distinct from the one discussed above.
We consider a qubit with initial state $
|\psi(0)\rangle =  |0\rangle, 
$
with Hamiltonian
$
H = \frac{\omega_0}{2}\,\sigma_z + \Omega \sigma_x.
$
Projectors on the $\sigma_z$ basis,
$
P_{0} = |0\rangle\langle0|~; ~ P_{1}=|1\rangle\langle1|
$.
Numerically we study this case, and we take the same parameter values as in case I, so the decay rates parameters will follow the same results as in Fig.\ref{Fig:Gamma_Area_KCV}(a) and therefore the positive and negative area of it will also be the same as Fig.\ref{Fig:Gamma_Area_KCV}(b), Now the KCC violation results $\text{viol}(t_1,t_2)$ with respect to cutoff frequency $\omega_c$ is shown in Fig. \ref{Fig:model2_KCV}.
And we can see that the KC violation follows the same nature with respect to the positive and negative area of the decay rates. 

In the non-Markovian regime, where KCC violation is governed by the behaviour of the decay rates, we proceed to establish analytical connections between different non-Markovianity measures following the RHP~\cite{Rivas_2014_RPM} and BLP~\cite{BLP_PRA_2010} criteria.

\section{Violation of Kolmogorov Consistency condition and the RHP measure of Non-Markovianity}
\label{KC_RHP}
The Rivas–Huelga–Plenio (RHP) measure quantifies non-Markovianity by capturing the breakdown of complete-positive (CP) divisibility in a quantum dynamical map. Precisely, it becomes nonzero whenever the intermediate map between two times fails to remain completely positive. For a time-local master equation with canonical decay rates $\chi_k=\{\Gamma_{\varepsilon},\widetilde{\gamma}_{\varepsilon}\}$, as in Eq.~(\ref{Eq:dyn_equ}), the RHP measure of non-Markovianity reduces to the formulation of non-Markovianity measure based on negative decay rates introduced by Andersson \textit{et al.}~\cite{Anderson_PRA_2014}. The equivalence between these two approaches was demonstrated in Ref.~\cite{Rivas_2014_RPM}, yielding the expression for the RHP measure of non-Markovianity in a two-dimensional Hilbert space as
\begin{equation}
\mathcal{N}_{\mathrm{RHP}}(T)=\int_{0}^{T}\sum_{k}\max\{0,-\chi_{k}(t)\}\,dt.
\end{equation}
Hence, according to Eq.~(\ref{Eq:lambda}) it becomes
\begin{equation}
\mathcal{N}_{\mathrm{RHP}}(T)=\int_{0}^{T}\lambda^{-}(s)\,ds=N(T),
\end{equation}
which, for case I, directly gives
\begin{equation}
\mathrm{viol}_+(t_{1},t_{2})
= \tfrac{1}{2}\,e^{-\tfrac{1}{2}M(t_{2})}\,
e^{+\tfrac{1}{2}\mathcal{N}_{\mathrm{RHP}}(t_{2})}\,\mathcal{S}(t_{1},t_{2}),
\end{equation}
where $\mathcal{S}=\big|\sin(\omega_0 t_1)\sin(\omega_0(t_2-t_1))\big|$.
Thus, the KCC violation increases exponentially with the RHP measure of non-Markovianity, while being simultaneously suppressed by the positive decay contributions and the oscillatory nature of the coherent dynamics.
Thus, the loss of CP-divisibility directly manifests as the observed Kolmogorov-consistency violation.

\section{Violation of Kolmogorov Consistency condition and the BLP measure of Non-Markovianity}
\label{KC_BLP}
The non-Markovianity measure introduced by Breuer \textit{et al.}~\cite{BLP_PRL_2009,BLP_PRA_2010} characterizes memory effects in open quantum dynamics through the backflow of information from the environment to the system. In this framework, non-Markovian behavior is identified when the distinguishability between two quantum states --- quantified by a suitable distance measure such as the trace distance --- increases over time. For two states $\rho_1(t)$ and $\rho_2(t)$, a temporary growth in their trace distance signals a reversal of information flow, indicating the breaking of monotonicity characterizing non-Markovian evolution.

Let us consider two quantum states, $\rho_1(t)$ and $\rho_2(t)$, evolving under the dynamical equation given in Eq.~(\ref{Eq:dyn_equ}). The trace distance between them at time $t$ is defined as
\begin{equation}
    D(t)=\frac{1}{2}\text{Tr}|\rho_1(t)-\rho_2(t)|.
\end{equation}
This quantity serves as a measure of their distinguishability. For a purely Markovian evolution, $D(t)$ decreases monotonically in time, while any temporary increase in $D(t)$ signals non-Markovian dynamics. The Breuer–Laine–Piilo (BLP) measure of non-Markovianity is defined as 
\begin{equation}
    \mathcal{N}_{\text{BLP}}=\max_{\rho_1(0)}\int_{\sigma > 0} dt \sigma(t,\rho_1(0),\rho_2(0)),
\end{equation}
where $\sigma(t,\rho_1(0),\rho_2(0))=\frac{d}{dt}D(t)$ and $\rho_2(0)$ is chosen as the state orthogonal to $\rho_1(0)$, since the maximal information backflow occurs for initially orthogonal states~\cite{Piilo_2012}.

Let us consider the general orthogonal initial pair of states $\rho_1(0)=|\psi_1\rangle\langle \psi_1|$ and  $\rho_2(0)=|\psi_2\rangle\langle \psi_2|$, with
\begin{align}
    &|\psi_1\rangle = \cos\frac{\delta}{2}\,|0\rangle+e^{i\zeta}\sin\frac{\delta}{2}\,|1\rangle,\nonumber\\
    &|\psi_2\rangle = -e^{-i\zeta}\sin\frac{\delta}{2}\,|0\rangle+\cos\frac{\delta}{2}\,|1\rangle,
\end{align}
respectively, where $0\le \delta \le \pi$ and $0 \le \zeta < 2\pi$. For this case, after evolving under the master equation, at time $t$ the trace distance becomes
\begin{equation}
D(t)=\sqrt{e^{-G(0,t)}\sin^2 \delta+e^{-2G(0,t)}\cos^2 \delta},
\end{equation}
and the BLP measure of non-Markovianity takes the form
\begin{align}
&\mathcal{N}_{\mathrm{BLP}}(t)
=
\max\bigl\{\widetilde{N}_1(t),\, \widetilde{N}_2(t)\bigr\},\nonumber\\
&\text{with} ~~\widetilde{N}_1(t) = \frac{1}{2}\int_0^t \lambda^{-}(s)\,e^{-\frac{1}{2}G(0,s)}\,ds,\nonumber\\
&\quad \quad \;\widetilde{N}_2(t) = \int_0^t \lambda^{-}(s)\,e^{-G(0,s)}\,ds.
\end{align}

From this expression of $\mathcal{N}_{\mathrm{BLP}}$ we obtain
\begin{align}
&\tfrac{1}{2}\mathcal{S}(t_1,t_2)\Big(1 - \tfrac{1}{2}A^+(t_2) + \tfrac{1}{2}\mathcal{N}_{\mathrm{BLP}}(t_2)\Big)
\;\le\;
\mathrm{viol}_+(t_1,t_2)
\;\nonumber\\
&\phantom{ami che}\le\;
\tfrac{1}{2}\mathcal{S}(t_1,t_2)\Big(1 - \tfrac{1}{2}A^+(t_2) +2 \mathcal{N}_{\mathrm{BLP}}(t_2)\Big),
\end{align}
Thus, $\mathrm{viol}_+(t_1,t_2)$ is bounded from below and above by linear functions of the BLP measure of non-Markovianity with slopes between $\mathcal{S}/4$ and $\mathcal{S}$ (with the same offset $\tfrac{1}{2}\mathcal{S}(t_1,t_2)(1-\tfrac{1}{2}A^+(t_2))$). See App.~\ref{App:BLP} for the detailed calculation.



In the next sections, we investigate the relation between KCC violation and information-theoretic and thermodynamic quantities.

\section{Relation Between Kolmogorov Consistency Violation and Quantum Mutual Information}
\label{KC_MI}

KCC violation reveals non-classical temporal behaviour, while mutual information measures system-environment correlations. Connecting these two quantities allows us to interpret KCC violation from an information-theoretic viewpoint and to investigate whether it directly originates from correlation backflow in non-Markovian evolution.


The reduced dynamics of the system, given by the Lindblad master equation in Eq.~(\ref{Eq:Master_Eq1}), can be represented by a global unitary evolution of the system and the reservoir followed by tracing out the reservoir degrees of freedom. Hence, the composite system-reservoir state at time $t$ is given by $\rho_{SR}(t)=U_{SR}(t)\,\rho(0)\otimes\rho_R\,U_{SR}^\dagger(t)$, where $U_{SR}(t)$ is the joint unitary evolution operator. The quantum mutual information between system and the reservoir at time $t$ is
\begin{equation}
I_{S:R}(t) = S(\rho(t)) + S(\rho_R) - S(\rho_{SR}(t)),
\label{mutu_info}
\end{equation}
where $S(\rho) = -\mathrm{Tr}(\rho \ln \rho)$ is the von Neumann entropy of the system $\rho$.  In Eq.~(\ref{mutu_info}), we have taken the Born approximation, which assumes that the coupling between the system and the reservoir is sufficiently weak such that system–reservoir correlations remain negligible and the state of the reservoir is approximately unchanged throughout the evolution.
Since $\rho_{SR}(t)$ evolves unitarily, $S(\rho_{SR}(t))=S(\rho_{SR}(0))=S(\rho(0))+S(\rho_R)$.  Therefore, $I_{S:R}(t) = S(\rho(t)) - S(\rho(0))$.
If the initial state is a pure state, then $S(\rho(0))=0$. Hence,
$
I_{S:R}(t) = S(\rho(t)).
$
So, the mutual information equals the von Neumann entropy of the reduced system.

The reduced density matrix of a single-qubit state evolving under the Markovian master equation discussed previously is given in Eq.~(\ref{eq:state1}). The von Neumann entropy of $\rho(t)$ is given by
\[
S(\rho(t)) = f\!\left[\frac{1+r(t)}{2}\right],
\]
where $f[p]=-p\ln p-(1-p)\ln(1-p)$ is the binary entropy function and $r(t)$ is the Bloch length of the state. For the state $\rho(t)$, the Bloch length $r(t)$ takes the form
\[
r(t) = \sqrt{(2a(t)-1)^2 + 4|c(t)|^2}.
\]

\textit{Relation to KCC Violation:}
For the initial state $|+\rangle$ and the measurement in the $\sigma_x$ eigenbasis, the KC violation for the outcome $|+\rangle \langle +|$ is given by
\[
\mathrm{viol}_+(t_1,t_2)
= |c(t_2)|\,|\sin(\omega_0 t_1)\sin[\omega_0(t_2-t_1)]|.
\]
Here, $|c(t_2)| = \tfrac{1}{2} e^{-\tfrac{1}{2}G(0,t_2)}$ (see Eq.~(\ref{eq:a_&_c})).
Thus, the KCC violation is directly proportional to the coherence amplitude of the state for fixed $\omega_0$, $t_1$, and $t_2$.
By substituting $|c(t_2)|$ from the above relation into the Bloch length, one obtains
\[
r(t_2)^2 = (2a(t_2)-1)^2 +
4\!\left(\frac{\mathrm{viol}_+(t_1,t_2)}{|\sin(\omega_0 t_1)\sin[\omega_0(t_2-t_1)]|}\right)^{\!2}.
\]
Consequently, the mutual information can be written explicitly in terms of the KCC violation as
\begin{align}&
I_{S:R}(t)=f\!\left[\frac{1+r(t)}{2}\right] \\
&=f\!\left[\frac{1}{2}\left(1+\sqrt{(2a(t_2)-1)^2 +
4\frac{\mathrm{viol}_+(t_1,t_2)^2}
{\sin^2(\omega_0 t_1)\sin^2[\omega_0\Delta]}}\right)\right].\nonumber
\end{align}
We can quantify how much mutual information (correlation) is present for a measured KCC violation from this relation.
\\
\\
\textbf{Corollaries and Bounds:}

\paragraph*{(i) Upper bound.}
Since $|\sin|\le1$, it follows that
\[
\mathrm{viol}(t_1,t_2)\le |c(t_2)|.
\]
Hence,
\[
r(t_2)^2 \ge (2a(t_2)-1)^2 + 4\,\mathrm{viol}(t_1,t_2)^2,
\]
which yields a lower bound on the mutual information in terms of the KCC violation and population imbalance.

\paragraph*{(ii) Balanced population case.} 
If $a(t_2)=1/2$, then
\[
r(t_2)=2|c(t_2)| = 2\,\frac{\mathrm{viol}(t_1,t_2)}{|\sin(\omega_0 t_1)\sin[\omega_0(t_2-t_1)]|},
\]
and therefore,
\[
I_{S:E}(t_2) = f\!\left(\frac{1}{2}\left[1+
2\,\frac{\mathrm{viol}(t_1,t_2)}{|\sin(\omega_0 t_1)\sin[\omega_0(t_2-t_1)]|}\right]\right).
\]
For small $\mathrm{viol}$, expanding the binary entropy gives
\[
I_{S:E}(t_2) \approx \frac{2}{\ln 2}\left(\frac{\mathrm{viol}(t_1,t_2)}{|\sin(\omega_0 t_1)\sin[\omega_0(t_2-t_1)]|}\right)^{\!2} + O(\mathrm{viol}^4),
\]
showing a quadratic dependence of mutual information on the KCC violation.

\section{Analytical Relation Between the Fano Factor and Kolmogorov Consistency Violation}
\label{KC_Fano}
The Fano factor provides a compact measure of fluctuations in any counting process by comparing the variance of events to their mean, and is widely used to distinguish classical, memoryless dynamics from processes with temporal correlations~\cite{Fano_PR_1947,Gardiner_book_2004,NazarovBlanter2009,VanKampen2007}. In both Poissonian and binary (Bernoulli) processes an independent, memoryless sequence yields a Fano factor of one, so any deviation from unity directly reflects the presence of temporal correlations. In our binary setting, all multi-time statistics are fully determined by such correlations, and a violation of the Kolmogorov Consistency Condition indicates that they cannot arise from any classical memoryless model. The Fano factor thus serves as an operational indicator of this behaviour: deviations from unity quantify the degree of correlation and hence signal KCC violation. Establishing this connection links the mathematical notion of KCC violation to an experimentally accessible observable, providing a clear physical interpretation of non-Markovian behaviour in binary stochastic processes.

\noindent\textbf{Definition of the Fano factor.}
To quantify measurement fluctuations, consider repeated projective measurements in the $\sigma_x$ basis at a fixed time $t$. The probability of obtaining the ``$+$'' outcome is
\[
p_+(t) = \frac{1}{2} + \Re[c(t)] = \frac{1}{2} + \frac{1}{2}e^{-\tfrac{1}{2}G(0,t)}\cos(\omega_0 t).
\]
For the binary (Bernoulli) random variable $X$ taking value $1$ for the ``$+$'' outcome and $0$ for the ``$-$'' outcome, the Fano factor is defined as the ratio between the variance and the mean:
\[
F_X(t) = \frac{\mathrm{Var}(X)}{\mathbb{E}[X]} = \frac{p_+(1-p_+)}{p_+} = 1 - p_+(t).
\]
Substituting $p_+(t)$ gives
\[
F_X(t) = \frac{1}{2} - \frac{1}{2}e^{-\tfrac{1}{2}G(0,t)}\cos(\omega_0 t)
= \frac{1}{2} - |c(t)|\cos(\omega_0 t).
\]
Hence, the Fano factor is linearly dependent on the real part of the system coherence.

\vspace{1em}
\noindent\textbf{Analytical connection between Fano factor and KC violation.}
We know that for the Case I, the KC Violation is
\[
\mathrm{viol}(t_1,t_2) = |c(t_2)|\,\big|\sin(\omega_0 t_1)\,\sin[\omega_0(t_2-t_1)]\big|.
\]
Eliminating $|c(t_2)|$ from the above expressions, we find an exact analytical relation between the Fano factor at time $t_2$ and the KCC violation between $t_1$ and $t_2$:
\[
F_X(t_2) = \frac{1}{2}
- \frac{\mathrm{viol}(t_1,t_2)}{\big|\sin(\omega_0 t_1)\,\sin[\omega_0(t_2-t_1)]\big|}\,\cos(\omega_0 t_2),
\]
valid for all $\omega_0 t_1, \omega_0 (t_2-t_1)$ such that the sine factors are nonzero.

In this setting, the measurement outcomes come from a Bernoulli process, since a single projective measurement in a two-level system produces two possible outcomes.
Non-Markovian information backflow sustains off-diagonal terms in the density matrix, thereby increasing 
F toward unity and enhancing Kolmogorov-consistency violation. 
The above relation shows that both $\mathrm{viol}(t_1,t_2)$ and $F_X(t_2)$ are governed by the same physical quantity, namely the coherence magnitude $|c(t)|$. Since $|c(t)|$ depends exponentially on the integral of $(\Gamma(t)+\widetilde{\gamma}(t))$, any negative decay rate temporarily enhances the coherence magnitude and, consequently, both $\mathrm{viol}(t_1,t_2)$ and deviations of $F_X$ from $1/2$. Therefore, the appearance of finite KCC violation can serve as an operational indicator of non-Markovian memory backflow, even in the presence of a thermal (mixed) environment. In particular, for times $t_1$ and $t_2$ chosen such that $\sin(\omega_0 t_1)=\sin[\omega_0(t_2-t_1)]=1$, the relation simplifies to
\[
F_X(t_2) = \frac{1}{2} - \mathrm{viol}(t_1,t_2)\cos(\omega_0 t_2).
\]
Hence, stronger KC violations correspond to stronger non-classical fluctuations (smaller Fano factor), especially when $\cos(\omega_0 t_2)>0$.

\section{Kolmogorov Consistency Violation : Heat exchange and Entropy Production Rate}
\label{KC_Q_sigma}
To clarify the physical meaning of Kolmogorov-consistency violation in open quantum systems, it is natural to ask whether this form of temporal non-classicality is linked to thermodynamic quantities and irreversibility. In this context, heat exchange between the system and the reservoir plays a central role, as it captures the energy flow associated with dissipation during the open-system evolution. Entropy production, which quantifies irreversibility through coherence loss and information backflow, provides a complementary thermodynamic measure. Comparing both heat exchange and entropy production with KCC violation can therefore reveal whether these phenomena stem from a common underlying physical mechanism rather than being unrelated features of the dynamics.


The heat absorbed by the system from the environment during the evolution from $t=0$ to $t=t_2$ is defined as
$
    Q = \mathrm{Tr}\!\big[\rho(t_2) H\big] - \mathrm{Tr}\!\big[\rho(0) H\big].
$
We get,
\begin{align}
 Q &= \omega_0 \left( \rho_{00}(t_2) - \tfrac{1}{2} \right ) \nonumber \\
 &= \omega_0 \left[
        e^{-G(0,t_2)}\left(\tfrac{1}{2} + \int_0^{t_2} e^{G(0,u)} \Gamma(u)\,du \right) - \tfrac{1}{2}
    \right],
    \label{Eq:heat}
\end{align}
Now as our target to Combine Relation between $\text{viol}(t_1,t_2)$ and the heat $Q$,  we identify $e^{-G(0,t_2)}$ and
Substituting the violation into \eqref{Eq:heat} yields the exact identity
\begin{align}
     Q  = 
    \omega_0 \Bigg[
        &\left(\frac{2\,\mathrm{viol}(t_1,t_2)}{
        \sin(\omega_0 t_1)\,\sin(\omega_0(t_2-t_1))}\right)^2 \nonumber \\
        &\left( \tfrac{1}{2} + \int_0^{t_2} e^{G(0,u)} \Gamma(u)\,du \right)
        - \tfrac{1}{2}
    \Bigg].
    \label{Eq:Q_Viol}
\end{align}
Here $Q\leq 0$ is taken as the system releases heat into the environment.
Equation \eqref{Eq:Q_Viol} is an exact algebraic relation between the KC violation and the heat exchanged. However, it involves not only the violation but also the additional functional
$
    I[\Gamma,\widetilde{\gamma}] := \int_0^{t_2} e^{G(0,u)} \Gamma(u)\,du ,
$
which depends on the detailed time profile of $\Gamma(t)$ or the whole history of the rates. This means that in general there is no one-to-one mapping between KC violation and heat. But there can be some special cases, like
\\
  \textit{Pure excitation:} If $\widetilde{\gamma}(t)\equiv 0$, then
    This yields the explicit witness function
    \begin{equation}
        W(Q) = \frac{\mathcal{S}}{2}\sqrt{1 - \frac{Q}{\omega_0}}.
    \end{equation}

Nevertheless, in limiting regimes such as pure excitation or pure decay, one obtains closed-form algebraic relations. These regimes highlight how KC violation is directly sensitive to non-Markovian effects encoded in negative or time-dependent rates, while the heat reflects the net energy exchange with the bath. Thus, comparing both quantities provides a richer understanding of memory effects: the violation captures coherence preservation, while the heat characterizes population dynamics. Together, they form complementary witnesses of non-Markovianity.
\\
\\
Now using Spohn’s formula~\cite{rivas2012},
\begin{align}
\sigma(t) &= -\mathrm{Tr}\!\left[\mathcal{L}_t(\rho(t))\big(\ln\rho(t)-\ln\rho_{\mathrm{eq}}\big)\right] \nonumber \\
&= \frac{d}{dt}S(\rho(t))-\beta\,\dot Q(t),
\end{align}
with $S(\rho)=-\mathrm{Tr}[\rho\ln\rho]$ and $\dot Q(t)=\omega_0\dot{a}(t)$,
Defining $r(t)=\sqrt{(a-\tfrac12)^2+|c|^2}$  gives
\[
\frac{dS}{dt}
= -\frac{(a-\tfrac12)\dot a+\Re[c^*\dot c]}{r}
\ln\!\left(\frac{\tfrac12+r}{\tfrac12-r}\right),
\]
Thus the exact entropy-production rate is
\begin{align}
\sigma(t)
 =& -\frac{(a-\tfrac12)(\Gamma-\lambda a)-\tfrac14\lambda e^{-G(0,t)}}{r(t)}
\ln\!\left(\frac{\tfrac12+r(t)}{\tfrac12-r(t)}\right)
\nonumber \\
&-\beta\,\omega_0\big(\Gamma(t)-\lambda(t)a(t)\big).
\label{eq:sigma_exact}
\end{align}

\textit{Approximate relation in the weak-coherence regime.} In many realistic cases the system, after some damping, is near the maximally mixed point $a\approx \frac{1}{2}$ and coherences are small. Make the small-r expansion $(r\ll 1)$. Then one may use the quadratic expansion of von Neumann entropy around maximally mixed. For $r\ll 1$, the entropy expansion $S\simeq\ln 2-2r^2 + O(r^4)$ yields
$\frac{dS}{dt}\simeq\frac{\lambda}{2} e^{-G(0,t)}$. Expressing $e^{-G}$ via 
$\mathrm{viol}$ gives the approximate scaling
\begin{equation}
\sigma(t_2)\;\approx\;
2\,\lambda(t_2)\,
\frac{\mathrm{viol}(t_1,t_2)^2}
{\sin^2(\omega_0 t_1)\,\sin^2(\omega_0(t_2-t_1))}
\;-\;\beta\,\omega_0\,\dot a(t_2).
\label{Eq:approx_sigma}
\end{equation}
In the small-coherence / near-maximally mixed regime the relation approximately holds: $\sigma$ is the sum of a term proportional to $\frac{\lambda(t)\text{viol}^2}{S^2}$ (coherence-driven) and a population/heat term $-\beta \omega_0 \dot{a}$. If population changes are small, $\sigma$ scales approximately like $\text{viol}^2$. Act as a witness $\dot{a}\approx0$,
\begin{equation}
    W(\sigma,\lambda)=\frac{\mathcal{S}}{\sqrt{2}}\sqrt{\frac{Q}{\lambda}}.
\end{equation}
Both KCC violation and entropy production stem from the same dynamical origin: the integrated rate $G(0,t)$ governing coherence decay. However, while $\mathrm{viol}$ depends only on $\lambda(t)=\Gamma(t)+\widetilde{\gamma}(t)$, the entropy production rate is additionally sensitive to the individual population contributions $\Gamma(t)$ and $\widetilde{\gamma}(t)$ through $\dot a(t)$. Therefore, a closed functional dependence $\sigma=\sigma(\mathrm{viol})$ does not exist in general. Nevertheless, this shows that KCC violation is suppressed in regimes where entropy production is dominated by diagonal relaxation, and is enhanced when coherence backflow plays a significant role. This establishes a physically meaningful link: both quantities can respond to the strength and character of information flow in the system, offering complementary perspectives on the breakdown of classical stochastic descriptions in non-Markovian quantum processes.
Nevertheless, in regimes where $\dot a\simeq 0$ and $r\ll 1$, Eq.~\eqref{Eq:approx_sigma} shows that entropy production is dominated by a term \emph{quadratic} in the KCC violation, establishing a direct quantitative link between memory effects and dissipation.

\section{Relation Between the Kirkwood--Dirac Quasi-Distribution and Kolmogorov Consistency}
\label{KC_KD}

Since both KCC violation and KD negativity arise from the temporal interplay of
quantum coherence and measurement disturbance, it is natural to investigate how
these two indicators are quantitatively connected. In particular, we aim to
establish an exact algebraic relation showing that KCC violation can be fully
interpreted as the net nonclassical interference contained in the KD
quasi-distribution. This connection provides a direct route for certifying
memory-induced nonclassical statistics through KD-based diagnostics.

For two projective measurements performed at times $t_{1}<t_{2}$ with projectors 
$\{P_{a}\}$ and $\{P_{b}\}$ respectively, the time-separated Kirkwood--Dirac (KD)
quasi-distribution is defined as
\begin{equation}
\mathrm{KD}(b,a)
:= \mathrm{Tr}\!\left[ P_b(t_2)\,P_a(t_1)\,\rho_0 \right],
\label{KD_def}
\end{equation}
where $P_{a}(t_{1})$ and $P_{b}(t_{2})$ are Heisenberg-picture operators and
$\rho_{0}$ is the initial state.
The KD distribution may take negative or complex values but always satisfy KCC as evident from the marginalizations
\begin{align}
\sum_{a}\mathrm{KD}(b,a) 
= & \mathrm{Tr}\!\left[P_b(t_2)\rho_0\right]
= p_{b}(t_{2}), \nonumber \\
\sum_{b}\mathrm{KD}(b,a)
=& \mathrm{Tr}\!\left[P_a(t_1)\rho_0\right]
= p_{a}(t_{1}).
\end{align}

The classical joint distribution constructed from intermediate projective collapse and conditional evolution is
\begin{equation}
p_{\mathrm{cl}}(b,a)
= p_{a}(t_{1})\, p(b|a),
\end{equation}
where $p(b|a)$ denotes the conditional probability of obtaining outcome $b$ at
$t_{2}$ given outcome $a$ at $t_{1}$. The KCC requires that
\begin{equation}
p_{b}(t_{2}) 
= \sum_{a} p_{a}(t_{1})\, p(b|a),
\end{equation}
and the violation of this condition is quantified by
\begin{equation}
\mathrm{viol}_{b}
= \left|
p_{b}(t_{2})
-
\sum_{a}
p_{a}(t_{1})\,p(b|a)
\right|.
\label{KC_violation_def}
\end{equation}

Define the KD interference contribution
\begin{equation}
\mathcal{I}(a,b)
:= \mathrm{KD}(b,a)
-
p_{a}(t_{1})\,p(b|a).
\end{equation}
Using the marginal $\sum_a\mathrm{KD}(b,a)=p_b(t_2)$ we obtain the exact identity
\begin{equation}
\sum_{a}\mathcal{I}(a,b)
=
p_{b}(t_{2})
-
\sum_{a}
p_{a}(t_{1})\,p(b|a),
\end{equation}
such that the KCC violation may be written as
\begin{equation}
\mathrm{viol}_{b}
=
\left|\sum_{a}\mathcal{I}(a,b)\right|
=
\left|\sum_{a}
\left(
\mathrm{KD}(b,a)
-
p_{a}(t_{1})\, p(b|a)
\right)\right|
.
\label{Eq:KD_KC_relation}
\end{equation}
Equation~\eqref{Eq:KD_KC_relation} shows that the KCC violation equals the total
nonclassical (interference) contribution of the KD distribution for a fixed
final outcome.
Thus, any nonclassicality (negativity or complex nature) in $\mathrm{KD}(b,a)$
is sufficient to induce KCC violation.

For the Case I. we can write

\begin{equation}
\label{Eq:Fano_model}
\mathrm{viol}_{+}
=
\left|\sum_{a\in\{+,-\}}
\Big[
\mathrm{KD}(+,a)
-
p_{a}(t_{1})\,p(+|a)
\Big]\right|
.
\end{equation}
From Eqs.~\eqref{Eq:viol2} and \eqref{Eq:Fano_model}, we obtain a closed-form relation between the Kolmogorov-consistency violation and the integrated decay rates ($\Gamma(t)$ and $\widetilde{\gamma}(t)$). This relation explicitly reveals a direct connection between memory effects and the interference of the KD quasi-probability in our model.
\section{Relation Between Kolmogorov Consistency Violation and Leggett--Garg Inequalities}
\label{KC_LGI}
Kolmogorov consistency and Leggett--Garg inequalities represent two distinct
classicality conditions imposed on temporal statistics of a single quantum
system. While Kolmogorov consistency constrains the compatibility between
single-time and multi-time probability distributions, Leggett--Garg inequalities
test the joint assumptions of macrorealism and non-invasive measurability through
temporal correlation functions. It is therefore natural to ask how violations of
Kolmogorov consistency are related to, or differ from, violations of
Leggett--Garg inequalities within the same dynamical framework.

To address this question, we analyze both criteria for a two-level system
undergoing open-system dynamics and measured at different times.
As a concrete example, we consider the following choice of dichotomic observable.
We consider the case I and as our dichotomic observable we choose
$Q=\sigma_x,\quad Q|q\rangle = q\,|q\rangle,\;\; q=\pm 1.$
\\
\\
\textit{Two-Time Correlation Functions:}
Because $|+\rangle$ is an eigenstate of $Q$, a projective measurement at $t=0$ leaves the state unchanged. Therefore the Heisenberg and sequential definitions of the correlator coincide. One finds
\begin{align}
C(0,t_1) &= \langle Q(0)\,Q(t_1)\rangle
= e^{-\tfrac{1}{2}G(0,t_1)}\cos(\omega_0 t_1), \\
C(t_1,t_2) &= \langle Q(t_1)\,Q(t_2)\rangle
= e^{-\tfrac{1}{2}G(t_1,t_2)}\cos\!\big(\omega_0(t_2-t_1)\big), \\
C(0,t_2) &= \langle Q(0)\,Q(t_2)\rangle
= e^{-\tfrac{1}{2}G(0,t_2)}\cos(\omega_0 t_2).
\end{align}
The derivation of $C(0,t_1)$, $C(0,t_2)$ and $C(t_1,t_2)$ are discussed in Appendix~\ref{LGI}.

\textit{Kolmogorov Consistency Violation:}
Kolmogorov consistency requires that the marginal probability at $t_2$ equals the sum of joint probabilities conditioned on outcomes at $t_1$. In terms of two-time correlators, this reduces to the multiplicative constraint
$ C(0,t_2)=C(0,t_1)\,C(t_1,t_2).$
Its violation is therefore quantified by
\begin{equation}
\label{Eq:KC_violation}
\mathrm{viol}_{+}
=
\tfrac{1}{2}\,
\big|\,C(0,t_1)\,C(t_1,t_2)-C(0,t_2)\,\big|.
\end{equation}
Substituting the explicit expressions above yields the closed form
\[
\mathrm{viol}_{+}
=
\tfrac{1}{2}\,
e^{-\tfrac{1}{2}G(0,t_2)}
\big|\sin(\omega_0 t_1)\,\sin\!\big(\omega_0(t_2-t_1)\big)\big|.
\]
Equation~\eqref{Eq:KC_violation} shows that KCC violation directly measures the \emph{failure of correlator factorization}.

\textit{Connection to the Leggett--Garg Inequality:}
A standard Leggett--Garg (LG) combination for three measurement times $0<t_1<t_2$ is
\begin{equation}
\label{Eq:LGI}
K_3 = C(0,t_1)+C(t_1,t_2)-C(0,t_2),
\end{equation}
with the macrorealist bound $K_3\le 1$.

Using Eq.~\eqref{Eq:KC_violation}, we may rewrite Eq.~\eqref{Eq:LGI} as
\begin{equation}
    K_3=\mathcal{F}(t_1,t_2)\;\pm\;2\,\mathrm{viol}_{+},
\end{equation}
where $\mathcal{F}(t_1,t_2)=C(0,t_1)C(t_1,t_2)-C(0,t_2)$,
where the sign depends on $\operatorname{sgn}[\,C(0,t_1)C(t_1,t_2)-C(0,t_2)\,]$ and $|\mathcal{F}(t_1,t_2)| \leq 1$.
Hence the LGI decomposes into a factorized classical term plus a correction proportional to the KCC violation. 
This relation further implies that any set of probability distributions satisfying the Kolmogorov consistency (KCC) condition cannot violate the Leggett–Garg inequalities (LGIs), as discussed in previous studies~\cite{Asano_PS_2014,Emary_RPP_2014}.

\section{Conclusion}
\label{Con}

In this work, we have shown that violations of the Kolmogorov consistency condition
provide a powerful and unifying signature of temporal non-classicality in open
quantum systems. Beyond merely signalling departures from classical Markovian
behaviour, KCC violation has been demonstrated to encode rich information about
memory effects, coherence dynamics, and thermodynamic exchanges between the system
and its environment.

We have first derived an exact analytical expression for the KCC violation in a
dissipative qubit model, enabling a transparent and quantitative understanding of
how such violations depend on the underlying non-Markovian features of the
dynamics. By working directly with the exact master equation, we have
explicitly related the magnitude of KCC violation to the integrated decay rates,
revealing a clear competition between positive and negative rate contributions.
While positive decay rates lead to an exponential suppression of KCC violation,
temporarily negative rates — a hallmark of non-Markovian memory effects —
exponentially enhance its magnitude. This establishes KCC violation as a sensitive
probe of coherence backflow and information return from the environment.
Our numerical analysis has corroborated the analytical trends by explicitly
simulating two representative dynamical scenarios. In both cases, we find that
the total positive area of the decay rates consistently exceeds the negative
contributions. As a result, the KCC violation is progressively suppressed at
longer times, despite the presence of temporary negative-rate intervals.
These results confirm that the behaviour and magnitude of KCC
violation are governed by the strength and character of the system–environment
interaction, and in particular by the balance between decoherence and memory
effects.

We have further established explicit connections between KCC violation and
well-known measures of non-Markovianity, including the RHP and BLP quantifiers.
Beyond non-Markovianity, we have connected KCC violation to
information-theoretic and thermodynamic quantities, including quantum mutual
information, the Fano factor, heat exchange, and entropy production. In particular,
we have shown that a measured KCC violation places a quantitative lower bound on
the mutual information shared between the system and its environment, thereby
providing a direct operational interpretation of temporal non-classicality in
terms of information correlations. The relation with the Fano factor further
establishes KCC violation as an experimentally accessible signature, linking it to
fluctuation properties of binary stochastic processes.
In certain dynamical regimes, we have demonstrated that the exchanged heat and the
entropy production rate can act as thermodynamic witnesses of KCC violation. While
a one-to-one functional dependence does not exist in general, both quantities are
ultimately governed by the same coherence-decay mechanisms, clarifying the role of
dissipation and irreversibility in suppressing or enhancing temporal quantum
effects. 
Finally, we have uncovered explicit relationships between KCC violation,
Kirkwood--Dirac negativity, and Leggett--Garg inequality violations within the same
dissipative qubit model. Our analysis shows that these seemingly distinct notions
of temporal quantumness share a common physical origin: coherence revivals induced
by non-Markovian memory effects associated with temporarily negative decay rates.

Taken together, our results position the Kolmogorov consistency condition as a
powerful criterion for assessing classicality in quantum stochastic dynamics. We show that non-Markovian memory effects and thermodynamic information flows leave clear and quantifiable imprints on this consistency. Viewed from this perspective, KCC violation provides a unifying and operational
notion of temporal non-classicality that bridges dynamical, informational, and
thermodynamic signatures, offering a conceptually transparent route to
classicality assessment in open quantum systems.

\section{Acknowledgement}

This research is supported by the Ministry of Education, Singapore, under its Academic Research Fund Programme (T2EP50222-0038 and RG154/24). A. Ghoshal acknowledges the support from the Alexander von Humboldt Foundation. 

\bibliography{tur_cm}

 \appendix
\begin{widetext}

\section{Calculation of the Kolmogorov-consistency condition for a single-qubit system}
\label{app:1}
We consider the system as a single qubit whose state at time $t$ is represented by the density matrix
\begin{align}
\rho(t)=\begin{pmatrix} a(t) & c(t) \\ c^*(t) & 1-a(t) \end{pmatrix}.
\label{state}
\end{align}
The Hamiltonian of the system is $H_{\text{sys}}=\frac{\hbar}{2}\omega_0\sigma_z$ and the time-local master equation under which the system is evolving is given by
\begin{equation}
    \frac{d\rho(t)}{dt} = -\frac{i}{\hbar}\big[H_{\text{sys}}, \rho(t)\big] + \Gamma_{\varepsilon}(t)\,L_{\sigma_-}[\rho(t)]+ \widetilde{\gamma}_{\varepsilon}(t)\,
L_{\sigma_+}[\rho(t)].
    \label{Eq:Master_Eq}
\end{equation}
with $L_{A(\varepsilon)}[\rho(t)]
= A(\varepsilon)\rho(t)A^{\dagger}(\varepsilon)
-\tfrac{1}{2}\{A^{\dagger}(\varepsilon)A(\varepsilon),\rho(t)\}$.
From the unitary part of Eq.~\eqref{Eq:Master_Eq} we obtain $\big[-\frac{i}{\hbar}[H_{\text{sys}},\rho(t)]\big]_{00}=
0$ 
as the diagonal entries of $H_{\text{sys}}$ are $0$. Thus, the Hamiltonian evolution does not affect the population dynamics. Next, evaluating the dissipative contributions yields
\[
\big[\Gamma_\varepsilon(t)L_{\sigma_-}[\rho(t)]\big]_{00}=\Gamma_\varepsilon(t)(1-a(t)), \qquad
\big[\widetilde{\gamma}_{\varepsilon}(t)L_{\sigma_+}[\rho(t)]\big]_{00}=-\widetilde{\gamma}_{\varepsilon}(t)a(t).
\]
Combining these results, the equation of motion for $a(t)$ becomes
\[
\dot a(t)
=\Gamma_\varepsilon(t)-\lambda(t)a(t)
\]
with $\lambda(t):=\Gamma_{\varepsilon}(t)+\widetilde{\gamma}_{\varepsilon}(t)$.
Similarly, for the coherence term $c(t)$ we get 
\begin{equation*}
\dot c(t) = -\Big(i\omega_0 + \tfrac{1}{2}\lambda(t)\Big)c(t).
\label{eq:popcoh}
\end{equation*}
Defining $G(t_0,t):=\int_{t_0}^{t}\lambda(s)\,ds$,
we get
\begin{equation}
a(t)=e^{-G(t_0,t)}\Big(a(t_0)+\int_{t_0}^t e^{G(t_0,s)}\Gamma_\varepsilon(s)\,ds\Big),
\qquad
c(t)=c(t_0)e^{-i\omega_0 (t-t_0)}\,e^{-\tfrac12 G(t_0,t)}.
\label{Aeq:a_c}
\end{equation}

To examine the validity of the Kolmogorov consistency condition, we need to calculate the quantity $\text{viol}(t_1,t_2)
= \Big | \sum_{x_1} P(x_2,x_1) - P(x_2) \Big |$.
Here, $P(x_2,x_1)$ denotes the joint probability of obtaining the measurement outcome $x_1$ at time $t_1$ and $x_2$ at time $t_2$, while $P(x_2)$ represents the single-time probability of getting the outcome $x_2$ at $t_2$ when no prior measurement is performed at $t_1$. To evaluate these probabilities, we must specify a measurement basis.
We consider a general orthonormal measurement basis defined as
\[
|u_1\rangle=\cos\frac{\theta}{2}\,|0\rangle+e^{i\phi}\sin\frac{\theta}{2}\,|1\rangle,
\qquad |u_2\rangle=-e^{-i\phi}\sin\frac{\theta}{2}\,|0\rangle+\cos\frac{\theta}{2}\,|1\rangle,
\]
where $0\le \theta \le \pi$ and $0 \le \phi < 2\pi$. Let us now calculate the quantity $\text{viol}(t_1,t_2)$ for the measurement outcome corresponding to $|u_1\rangle$. To do this, we need to determine two kinds of probabilities: the single-time probabilities and the conditional probabilities. The single-time probabilities are obtained by letting the system evolve from its initial state up to the measurement time and then performing a projective measurement. The conditional probabilities are obtained by first measuring the system at an intermediate time, allowing the post-measurement state to evolve further, and then performing a second measurement. These probabilities together allow us to evaluate the Kolmogorov-consistency violation.\\

\noindent
\textbf{Single-time probabilities:}
Consider that the system evolves from the initial time $t=0$ to $t=t_i$ according to the master equation. At time $t_i$, we perform a projective measurement in the basis $\{|u_1\rangle,|u_2\rangle\}$. Upon measurement, the state collapses to either $|u_1\rangle\langle u_1|$ or $|u_2\rangle\langle u_2|$. The probability of obtaining the outcome $|u_1\rangle$ or $|u_2\rangle$ at time $t_i$ is given by
\begin{equation}
\label{Eq:prob}
    p_{u_1}(t_i)= \Tr[|u_1\rangle \langle u_1|\rho(t_i)]=\frac{1}{2}+\big(a(t_i)-\frac{1}{2}\big )\cos \theta
+ \sin \theta\,\Re\big[e^{i\phi}c(t_i)\big] \quad \text{and} \quad p_{u_2}(t_i)=1-p_{u_1}(t_i),
\end{equation}
with 
\begin{equation}
\label{eq:a_&_c}
a(t_i)=e^{-G(0,t_i)}\Big(a(0)+\int_{0}^{t_i} e^{G(0,s)}\Gamma_\varepsilon(s)\,ds\Big),
\qquad
c(t_i)=c(0)e^{-i\omega_0 t_i}\,e^{-\tfrac12 G(0,t_i)}.
\end{equation}
\textbf{Post-measurement states at \(t_1\):}
If at time $t_1$ a projective measurement is performed in the basis $\{|u_1\rangle,|u_2\rangle\}$, the system state collapses to one of the corresponding projectors depending on the measurement outcome. If the outcome is $|u_1\rangle$, the post-measurement state at time $t_1$ becomes $\rho_{u_1}(t_1)=|u_1\rangle\langle u_1|$, whose matrix elements in the computational basis are
\[
a_{u_1}(t_1)=\cos^2\!\frac{\theta}{2}=\frac{1+\cos\theta}{2},\qquad
c_{u_1}(t_1)=\frac{1}{2}e^{-i\phi}\sin\theta.
\]
Similarly, if the outcome corresponds to 
$|u_2\rangle$, the post-measurement state $\rho_{u_2}(t_1)=|u_2\rangle\langle u_2|$ has the matrix elements
\[
a_{u_2}(t_1)=\sin^2\!\frac{\theta}{2}=\frac{1-\cos\theta}{2},\qquad
c_{u_2}(t_1)=-\frac{1}{2}e^{-i\phi}\sin\theta.
\]
\textbf{Evolution from \(t_1\) to \(t_2=t_1+\Delta\):}
After the measurement at time $t_1$, the post-measurement state $\rho_{u_i}(t_1)$ 
$(i=1,2)$ evolves under the same time-local master equation given in Eq.~(\ref{Eq:Master_Eq}). The corresponding equations of motion for the matrix elements are provided in Eq.~(\ref{Aeq:a_c}). Therefore, the evolved populations and coherences at time $t_2$ can be written as
\[
a_{u_i}(t_2)
= e^{-G(t_1,t_2)}\,\Big [a_{u_i}(t_1)
\;+\;\!\int_{t_1}^{t_2} e^{G(t_1,s)} \Gamma_\varepsilon(s)\,ds\Big ],
\qquad i=1,2.
\]
and
\[
c_{u_i}(t_2)=c_{u_i}(t_1)\,e^{-i\omega_0\Delta-\tfrac{1}{2}G(t_1,t_2)}.
\]
\textbf{Conditional probabilities:}
Using the general expression for the single-time measurement probability given in Eq.~(\ref{Eq:prob}),   
the conditional probabilities of obtaining the outcome $|u_1\rangle$ at time $t_2$, given that the outcome at $t_1$ was either $|u_1\rangle$ or $|u_2\rangle$, are expressed as
\[
\begin{aligned}
p_{u_1}(t_2\!\mid\!u_i)
&= \frac{1}{2} + \big(a_{u_i}(t_2)-\frac{1}{2}\big)\cos\theta
+\sin\theta\;\Re\!\big[e^{i\phi}c_{u_i}(t_2)\big].
\end{aligned}
\]
\textbf{Kolmogorov--consistency violation:} To check the validity of Kolmogorov consistency condition, we evaluate the quantity $\text{viol}(t_1,t_2)$. In this case, for the outcome $|u_1\rangle$ it is defined as
\begin{equation}
    \text{viol}_{u_1}(t_1,t_2)=\big |p_{u_1}(t_2)- \Pi_{u_1}(t_2) \big |, 
\end{equation}
where $\Pi_{u_1}$ is the classical mixture defined as
\begin{equation}
    \Pi_{u_1}(t_2) := p_{u_1}(t_1)\,p_{u_1}(t_2\!\mid\!u_1) + p_{u_2}(t_1)\,p_{u_1}(t_2\!\mid\!u_2).
\end{equation}
Substituting the explicit forms of the single-time and conditional probabilities, and grouping terms according to diagonal (population) and off-diagonal (coherence) contributions, we can express the violation in a compact form given by
\[
\mathrm{viol}_{u_1}(t_1,t_2)
= \big|\,\mathcal{P}(t_1,t_2)\;+\;\mathcal{C}(t_1,t_2)\,\big|,
\]
where
\[
\begin{aligned}
\mathcal{P}(t_1,t_2)
&= \cos\theta\;\Big\{\,\big(a(t_2)-\tfrac{1}{2}\big)
- p_{u_1}(t_1)\big(a_{u_1}(t_2)-\tfrac{1}{2}\big) - p_{u_2}(t_1)\big(a_{u_2}(t_2)-\tfrac{1}{2}\big)\Big\},\\[6pt]
\mathcal{C}(t_1,t_2)
&= \sin\theta\;\Big\{\Re\big[e^{i\phi}c(t_2)\big ] - p_{u_1}(t_1)\,\Re\!\big[e^{i\phi}c_{u_1}(t_2)\big]
- p_{u_2}(t_1)\,\Re\!\big[e^{i\phi}c_{u_2}(t_2)\big] \Big\}.
\end{aligned}
\]
Similarly, we can calculate the KC violation term for the outcome $|u_2\rangle$. The presence of a nonzero $\text{viol}(t_1, t_2)$ thus signifies a breakdown of Kolmogorov consistency, reflecting the influence of non-classicality in the dynamics.
We now consider a few special cases to analyze the KC violation in greater detail.

\subsection{Case I}

We consider the system to be initially prepared in the state $\rho(0)=|\psi(0)\rangle\langle \psi(0)|$, with
\[
|\psi(0)\rangle = |+\rangle = \tfrac{1}{\sqrt{2}}\big(|0\rangle + |1\rangle\big),
\]
and the measurements are performed in the basis $\{|+\rangle, |-\rangle\}$. This scenario corresponds to the parameters $a(0)=c(0)=\frac{1}{2}$, $\theta=\frac{\pi}{2}$, and $\phi=0$. For this case, the single-time and conditional probabilities associated with obtaining the outcome $|+\rangle$ are given by
\begin{eqnarray}
&&p_+(t_i) = \tfrac12 + \Re[c(t_i)]
= \tfrac12 + \tfrac12 e^{-\tfrac12 G(0,t_i)}\cos(\omega_0 t_i), \nonumber\\
&&p_{+}(t_2\!\mid\!\pm)= \frac{1}{2}
+\Re\!\big[c_{\pm}(t_2)\big] =\frac{1}{2}\pm \frac{1}{2}e^{-\tfrac{1}{2}G(t_1,t_2)}\cos(\omega_0\Delta),
\label{eq:pplus_expr}
\end{eqnarray}
The Kolmogorov consistency condition for outcome $+$ demands
\[
p_+(t_2)\overset{?}{=} p_+(t_1)\,p_+(t_2|+) + p_-(t_1)\,p_+(t_2|-).
\]
Let us now define
\[
A:=\tfrac12 e^{-\tfrac12 G(0,t_1)}\cos(\omega_0 t_1),\quad
B:=\tfrac12 e^{-\tfrac12 G(t_1,t_2)}\cos(\omega_0\Delta),\quad
C:=\tfrac12 e^{-\tfrac12 G(0,t_2)}\cos(\omega_0 t_2),
\]
so that $p_\pm(t_1)=\tfrac12\pm A$, $p_+(t_2|\pm)=\tfrac12 \pm B$, $p_+(t_2)=\tfrac12+C$, and a short algebra yields
\[
p_+(t_1)p_+(t_2|+) + p_-(t_1)p_+(t_2|-) = \tfrac12 + 2AB.
\]
Hence, the Kolmogorov-consistency violation for outcome $|+\rangle \langle +|$ is
\[
\mathrm{viol}_+(t_1,t_2)=\big|2AB-C\big|.
\]
Using $\cos(\omega_0 t_2)=\cos(\omega_0 t_1)\cos(\omega_0\Delta)-\sin(\omega_0 t_1)\sin(\omega_0\Delta)$ and $G(0,t_2)=G(0,t_1)+G(t_1,t_2)$ we get
\[
2AB-C = \tfrac12 e^{-\tfrac12 G(0,t_2)}\sin(\omega_0 t_1)\sin(\omega_0\Delta).
\]
Therefore,
\begin{align}
\mathrm{viol}_+(t_1,t_2)
&= \tfrac12\,\exp\!\Big(-\tfrac12 G(0,t_2)\Big)
\big|\sin(\omega_0 t_1)\sin(\omega_0(t_2-t_1))\big|\nonumber\\
&= \tfrac12\,\exp\!\Big(-\tfrac12\int_0^{t_2}\lambda(s)\,ds\Big)\,
\big|\sin(\omega_0 t_1)\sin(\omega_0(t_2-t_1))\big|.
\label{eq:viol1}
\end{align}
Substituting the expression for the time-dependent decay rate $\lambda(s)$, we finally obtain
\begin{equation}
\mathrm{viol}_+(t_1,t_2)
= \tfrac12\,\exp\!\Big(-\tfrac12\int_0^{t_2}\bigg[\Gamma_{\varepsilon}(s)+\widetilde{\gamma}_{\varepsilon}(s)\bigg]\,ds\Big)\,
\big|\sin(\omega_0 t_1)\sin(\omega_0(t_2-t_1))\big|.
\label{eq:viol_g}
\end{equation}

Since the setup is symmetric with respect to the measurement outcomes $|+\rangle$ and $|-\rangle$, the same result holds for both cases, i.e., $\text{viol}_+(t_1,t_2)=\text{viol}_-(t_1,t_2)$.  
For notational simplicity, we therefore drop the outcome subscript and
denote the violation as $\mathrm{viol}(t_1,t_2)$. To reveal the effects of positive and negative decay-rate contributions, we decompose any real function $x(t)$ into its positive and negative parts, $x(t)=x^+(t)-x^-(t)$ with $x^\pm(t)\ge0$. Applying this to the total decay rate $\lambda(t)=\Gamma_\varepsilon(t)+\widetilde{\gamma}_{\varepsilon}(t)$ with $\lambda^{\pm}(t)=\Gamma^{\pm}_{\varepsilon}(t)+\widetilde{\gamma}_{\varepsilon}^{\pm}(t)$, we define
\[
\lambda(t)=\lambda^+(t)-\lambda^-(t),\qquad
M(t_2):=\int_0^{t_2}\lambda^+(s)\,ds,\qquad
N(t_2):=\int_0^{t_2}\lambda^-(s)\,ds,
\]
so that $\int_0^{t_2}\lambda(s)\,ds=M(t_2)-N(t_2)$. Substituting into \eqref{eq:viol_g} yields the factorized form
\begin{equation}
    \label{eq:viol_MN}
    \mathrm{viol}_+(t_1,t_2) = \tfrac{1}{2}\,e^{-\tfrac{1}{2}\big(M(t_2)-N(t_2)\big)}\big|\sin(\omega_0 t_1)\sin(\omega_0(t_2-t_1))\big|.
\end{equation}
The negative parts of the decay rates, represented by $\lambda^-(t)$, correspond to temporary reversals in the direction of information flow between the system and its environment. These intervals signal the backflow of information, where coherence or correlations that were previously lost to the environment are partially restored to the system. Such behavior is a defining feature of non-Markovian dynamics, marking deviations from purely memoryless evolution. Hence,
\begin{equation}
\mathrm{viol}_+(t_1,t_2)
=
\underbrace{\tfrac12\,e^{-\tfrac12 M(t_2)}\big|\sin(\omega_0 t_1)\sin(\omega_0(t_2-t_1))\big|}_{\text{positive-rate damping}}
\;\times\;
\underbrace{e^{+\tfrac12 N(t_2)}}_{\text{negative-rate (non-Markovian) enhancement}}.
\label{eq:viol_factor_g}
\end{equation}

\section{Relation between BLP measure of non-Markovianity and KC violation}
\label{App:BLP}

Let us consider two quantum states, $\rho_1(t)$ and $\rho_2(t)$, evolving under the dynamical equation given in Eq.~(\ref{Eq:dyn_equ}). The trace distance between them at time $t$ is defined as
\begin{equation}
    D(t)=\frac{1}{2}\text{Tr}|\rho_1(t)-\rho_2(t)|.
\end{equation}
This quantity serves as a measure of their distinguishability. For a purely Markovian evolution, $D(t)$ decreases monotonically in time, while any temporary increase in $D(t)$ signals non-Markovian dynamics. The Breuer–Laine–Piilo (BLP) measure of non-Markovianity is defined as 
\begin{equation}
    \label{Eq:BLP}\mathcal{N}_{\text{BLP}}=\max_{\rho_1(0)}\int_{\sigma > 0} dt \sigma(t,\rho_1(0),\rho_2(0)),
\end{equation}
where $\sigma(t,\rho_1(0),\rho_2(0))=\frac{d}{dt}D(t)$ and $\rho_2(0)$ is chosen as the state orthogonal to $\rho_1(0)$, since the maximal information backflow occurs for initially orthogonal states.

Let us consider the general orthogonal initial pair of states 
\begin{align}
    &\rho_1(0)=|\psi_1\rangle\langle \psi_1| \quad \text{with} \quad |\psi_1\rangle = \cos\frac{\delta}{2}\,|0\rangle+e^{i\zeta}\sin\frac{\delta}{2}\,|1\rangle,\nonumber\\
    &\rho_2(0)=|\psi_2\rangle\langle \psi_2| \quad \text{with} \quad |\psi_2\rangle = -e^{-i\zeta}\sin\frac{\delta}{2}\,|0\rangle+\cos\frac{\delta}{2}\,|1\rangle,
\end{align}
where $0\le \delta \le \pi$ and $0 \le \zeta < 2\pi$. Under the dynamics governed by the time-local master equation, the evolved states at time $t$ are
\[
\rho_1(t)=\begin{pmatrix} \cos^2 \frac{\delta}{2}e^{-G(0,t)}+Q(0,t) & \frac{1}{2} z(t)\\ \frac{1}{2} z^*(t)& \sin^2 \frac{\delta}{2}e^{-G(0,t)}+Q(0,t)\end{pmatrix},
\]
\[
\rho_2(t)=\begin{pmatrix} \sin^2 \frac{\delta}{2}e^{-G(0,t)}+Q(0,t) & -\frac{1}{2}z(t)\\ -\frac{1}{2}z^*(t)& \cos^2 \frac{\delta}{2}e^{-G(0,t)}+Q(0,t) \end{pmatrix},
\]
where $z(t)=\sin \delta e^{-i(\zeta+\omega_0t)}\,e^{-\tfrac12G(0,t)}$ and $Q(0,t)=e^{-G(0,t)}\int_{0}^{t} e^{G(0,s)}\Gamma_\varepsilon(s)\,ds$. The trace distance between these two states, which quantifies their distinguishability, is then
\begin{equation}
\label{eq:D}
D(t)=\frac{1}{2}\text{Tr}|\rho_1(t)-\rho_2(t)|=\sqrt{|z(t)|^2+e^{-2G(0,t)}\cos^2 \delta}=\sqrt{e^{-G(0,t)}\sin^2 \delta+e^{-2G(0,t)}\cos^2 \delta}.
\end{equation}
Differentiating $D(t)$ with respect to time gives
\begin{equation}
    \sigma(t,\delta)=\frac{d}{dt}D(t)=-\frac{\lambda(t)}{2D(t)}(e^{-G(0,t)}\sin^2 \delta+2e^{-2G(0,t)}\cos^2 \delta).
\end{equation}
Here, the dependence of the evolved states on the initial conditions reduces entirely to the single parameter $\delta$. It follows that $\sigma(t,\delta)>0$ only when $\lambda(t)<0$, that is, during intervals where information flows back from the environment to the system. Denoting $\lambda^-(t)=\max\{-\lambda(t),0\}$, we can write
\begin{equation}
 \text{for} \;\;\sigma(t,\delta)>0,\quad   \sigma(t,\delta)=\frac{\lambda^-(t)}{2D(t)}(e^{-G(0,t)}\sin^2 \delta+2e^{-2G(0,t)}\cos^2 \delta).
\end{equation}
The BLP measure of non-Markovianity integrates these positive contributions of $\sigma(t,\delta)$ over time and maximizes over all initial pairs of states as given in Eq.~(\ref{Eq:BLP}). Maximising over $\delta$ for a fixed $t$ gives two regimes:
\begin{align}
    &\text{For}\; e^{-G(0,t)}<\frac{1}{4},\quad \delta_{\text{max}}=\frac{\pi}{2},\quad \;\ \sigma(t,\frac{1}{2})=\frac{1}{2}\lambda^-(t)e^{-\frac12G(0,t)}\nonumber\\
    & \text{For}\; e^{-G(0,t)}>\frac{1}{4},\quad \delta_{\text{max}}=0, \quad \quad\sigma(t,0)=\lambda^-(t)e^{-G(0,t)}.
\end{align}
This shows that the optimal initial states for the BLP measure change depending on the effective decoherence captured by $G(0,t)$. Hence, the BLP measure of non-Markovianity up to time $t$ can be expressed as
\begin{equation}
\label{BLP}
\mathcal{N}_{\mathrm{BLP}}(t)
=
\max\bigl\{\widetilde{N}_1(t),\, \widetilde{N}_2(t)\bigr\}, ~~~ \text{with} ~~\widetilde{N}_1(t) = \frac{1}{2}\int_0^t \lambda^{-}(s)\,e^{-\frac{1}{2}G(0,s)}\,ds,
~~\widetilde{N}_2(t) = \int_0^t \lambda^{-}(s)\,e^{-G(0,s)}\,ds.
\end{equation}
Let us now establish the relation between the KC violation term given in Eq.~(\ref{eq:viol_g}) and the BLP measure of non-Markovianity $\mathcal{N}_{\mathrm{BLP}}$.
For $\delta_{\max}=\frac{\pi}{2}$, $\dot D(t)=-\tfrac12\lambda(t)D(t)$. Integrating from $0$ to $t_2$, with $D(0)=1$, gives
\begin{align}
\label{eq:D_integral}
D(t_2)=1-\frac{1}{2}\int_0^{t_2}\lambda(s)\,e^{-\tfrac12G(0,s)}\,ds
= 1-\frac{1}{2}A^\dagger(t_2)
+\widetilde{N}_1(t_2),
\end{align}
where $A^+(t_2) := \int_0^{t_2} \lambda^+(s)\,e^{-\frac{1}{2}G(0,s)}\,ds $. The KC violation term for case I reads
\begin{align}
    \mathrm{viol}_+(t_1,t_2)
= \tfrac12\,\exp\!\Big(-\tfrac12 G(0,t_2)\Big)
\big|\sin(\omega_0 t_1)\sin(\omega_0(t_2-t_1))\big|,
\end{align}
which for $\delta_{\max}=\frac{\pi}{2}$ gives
\begin{align}
    \mathrm{viol}_+(t_1,t_2)
&= \tfrac12\,D(t_2)
\big|\sin(\omega_0 t_1)\sin(\omega_0(t_2-t_1))\big|.
\end{align}
Substituting $D(t_2)$ here, the Kolmogorov consistency violation becomes
\begin{equation}
\mathrm{viol}_+(t_1,t_2)
=
\frac{1}{2}
\left|
\sin(\omega_0 t_1)\,
\sin(\omega_0 (t_2 - t_1))
\right|\,
\bigl[
1 - \tfrac{1}{2}A^+(t_2) + \widetilde{N}_1(t_2)
\bigr].
\end{equation}
We can relate $\widetilde{N}_1(t)$ and $\widetilde{N}_2(t)$ analytically because the functions $e^{-\tfrac12 G(0,t)}$ and $e^{-G(0,t)}$ satisfy,
\begin{equation}
0 < e^{-G(0,t)} \le e^{-\frac{1}{2}G(0,t)} \le 1 \qquad (\text{for } G(0,t)\ge0),
\end{equation}
Hence, integrating over time gives
\begin{equation}
\frac{1}{2}\widetilde{N}_2(t) \le \,\widetilde{N}_1(t).
\end{equation}
Therefore, from Eq.~(\ref{BLP}) we get
\begin{equation}
\frac{1}{2}\,\mathcal{N}_{\mathrm{BLP}}(t)
\le
\widetilde{N}_1(t)
\le
2\mathcal{N}_{\mathrm{BLP}}(t).
\end{equation}
So, we obtain
\begin{equation}
\tfrac{1}{2}\mathcal{S}(t_1,t_2)\Big(1 - \tfrac{1}{2}A^+(t_2) + \tfrac{1}{2}\mathcal{N}_{\mathrm{BLP}}(t_2)\Big)
\;\le\;
\mathrm{viol}_+(t_1,t_2)
\;\le\;
\tfrac{1}{2}\mathcal{S}(t_1,t_2)\Big(1 - \tfrac{1}{2}A^+(t_2) +2 \mathcal{N}_{\mathrm{BLP}}(t_2)\Big),
\end{equation}
where $\mathcal{S}=\big|\sin(\omega_0 t_1)\sin(\omega_0(t_2-t_1))\big|$. Thus, $\mathrm{viol}_+(t_1,t_2)$ is bounded from below and above by linear functions of the BLP measure of non-Markovianity with slopes between $\mathcal{S}/4$ and $\mathcal{S}$ (with the same offset $\tfrac{1}{2}\mathcal{S}(t_1,t_2)(1-\tfrac{1}{2}A^+(t_2))$).



\section{Relation between Kolmogorov Consistency Violation and Heat Exchange}

In this section we analyze the connection between the violation of the Kolmogorov consistency (KC) condition and the heat exchanged with the environment for a two-level system undergoing non-Markovian open system dynamics. Here we consider case-I. For this we already derived in \ref{Aeq:a_c} the solutions are
\begin{align}
    \rho_{01}(t) &= \frac{1}{2} e^{-i\omega_0 t}\,e^{-\tfrac{1}{2} G(0,t)}, \\
    \rho_{00}(t) &= e^{-G(0,t)}\left(\tfrac{1}{2} + \int_0^t e^{G(0,u)} \Gamma(u)\,du\right).
\end{align}
$G(0,t_2)=\int_{0}^{t_2}\lambda(s)ds$.
Now for initial state as $\rho(0)=|+\rangle\langle+|$, with $|+\rangle = (|0\rangle+|1\rangle)/\sqrt{2}$, and projective measurements in the $\sigma_x$ basis.
The violation of KC at two times $t_1 < t_2$ is found to be
\begin{equation}
    \mathrm{viol}(t_1,t_2) 
    = \frac{1}{2}\, e^{-\tfrac{1}{2} G(0,t_2)} 
      \left|\sin(\omega_0 t_1)\, \sin\big(\omega_0 (t_2-t_1)\big)\right|.
    \label{Aeq:KCviol2}
\end{equation}

The heat absorbed from the environment between $t=0$ and $t=t_2$ is defined as
\begin{equation}
    Q = \mathrm{Tr}\!\big[\rho(t_2) H\big] - \mathrm{Tr}\!\big[\rho(0) H\big].
\end{equation}
Since $\langle H \rangle_t = \omega_0\left(\rho_{00}(t)-\tfrac{1}{2}\right)$, and $\rho_{00}(0)=1/2$, we obtain
\begin{equation}
    Q = \omega_0 \left( \rho_{00}(t_2) - \tfrac{1}{2} \right).
\end{equation}
Substituting the exact solution for $\rho_{00}(t)$ gives
\begin{equation}
    Q = \omega_0 \left[
        e^{-G(0,t_2)}\left(\tfrac{1}{2} + \int_0^{t_2} e^{G(0,u)} \Gamma(u)\,du \right) - \tfrac{1}{2}
    \right].
    \label{eq:Heat}
\end{equation}

Now as our target to Combine Relation between $\text{viol}(t_1,t_2)$ and the heat $Q$, 
from Eq.~\ref{Aeq:KCviol2} we identify
\begin{equation}
    e^{-G(0,t_2)} = \left(\frac{2\,\mathrm{viol}(t_1,t_2)}{
    \sin(\omega_0 t_1)\,\sin(\omega_0(t_2-t_1))}\right)^2 .
    \label{eq:expG}
\end{equation}
Substituting \eqref{eq:expG} into \eqref{eq:Heat} yields the exact identity
\begin{equation}
    Q = \omega_0 \left[
        \left(\frac{2\,\mathrm{viol}(t_1,t_2)}{
        \sin(\omega_0 t_1)\,\sin(\omega_0(t_2-t_1))}\right)^2
        \left( \tfrac{1}{2} + \int_0^{t_2} e^{G(0,u)} \Gamma(u)\,du \right)
        - \tfrac{1}{2}
    \right].
    \label{eq:Q_vs_Viol}
\end{equation}
Equation \eqref{eq:Q_vs_Viol} is an \emph{exact algebraic relation} between the KC violation and the heat exchanged. However, it involves not only the violation but also the additional functional
\begin{equation}
    I[\Gamma,\widetilde{\gamma}] := \int_0^{t_2} e^{G(0,u)} \Gamma(u)\,du ,
\end{equation}
which depends on the detailed time profile of $\Gamma(t)$ or the whole history of the rates. This means that in general there is no one-to-one mapping between KC violation and heat. But there can be some special cases, like
\begin{itemize}
\item \textbf{Symmetric Rates:} $\Gamma(t)=\widetilde{\gamma}(t)=\frac{1}{2}\lambda(t)$.
\\
\\
The $\lambda(t)=2\Gamma(t)$ and $\Gamma=\widetilde{\gamma}$ with initial $a(0)=\frac{1}{2}$ one finds $a(t)=\frac{1}{2}$ for all $t$ (because the ODE $\dot{a}=\Gamma-\lambda a=\Gamma-2\Gamma$ has $a=\frac{1}{2}$ solution with that initial condition). Therefore $Q=0$ identically, while $\text{viol}=\frac{1}{2}e^{-G/2}S$ may be non-zero. 

    \item \textbf{Pure excitation:} If $\widetilde{\gamma}(t)\equiv 0$, then
    \[
        Q = \omega_0\Big(1 - e^{-G(0,t_2)}\Big)
        = \omega_0\left(1 - \Big(\tfrac{2\,\mathrm{viol}(t_1,t_2)}{S}\Big)^2\right),
    \]
    where $S = \sin(\omega_0 t_1)\sin(\omega_0(t_2-t_1))$.
    This yields the explicit relation
    \begin{equation}
        \mathrm{viol}(t_1,t_2) = \frac{S}{2}\sqrt{1 - \frac{Q}{\omega_0}}.
    \end{equation}

    \item \textbf{Pure decay:} If $\Gamma(t)\equiv 0$, then
    \[
        Q = \frac{\omega_0}{2}\big(e^{-G(0,t_2)} - 1\big),
    \]
    and thus
    \begin{equation}
        \mathrm{viol}(t_1,t_2) = \frac{S}{2}\sqrt{1 + \frac{2Q}{\omega_0}}.
    \end{equation}
    Here $Q\leq 0$ as the system releases heat into the environment.
\end{itemize}


\section{Kolmogorov Consistency Violation and Entropy Production}

\textbf{KC violation.}
For the case-I, the initial state $\ket{+}=(\ket{0}+\ket{1})/\sqrt{2}$, Hamiltonian 
$H=(\omega_0/2)\,\sigma_z$, and projective measurements in the $\sigma_x$ basis
at times $t_1<t_2$, the Kolmogorov consistency (KC) violation is
\begin{equation}
\label{eq:KCviol}
\mathrm{viol}(t_1,t_2)
= \tfrac{1}{2}\,e^{-\tfrac{1}{2}G(0,t_2)}
\Big|\sin(\omega_0 t_1)\,\sin\!\big(\omega_0 (t_2-t_1)\big)\Big|,
\end{equation}
with $G(0,t)=\int_0^t\lambda(s)\,ds$ and $\lambda(t)=\Gamma(t)+\widetilde{\gamma}(t)$.

\textbf{Entropy production rate.}
Using Spohn’s formula,
\begin{equation}
\sigma(t) = -\mathrm{Tr}\!\left[\mathcal{L}_t(\rho(t))\big(\ln\rho(t)-\ln\rho_{\mathrm{eq}}\big)\right]
= \frac{d}{dt}S(\rho(t))-\beta\,\dot Q(t),
\end{equation}
with $S(\rho)=-\mathrm{Tr}[\rho\ln\rho]$ and $\dot Q(t)=\omega_0\dot{a}(t)$,
we write the state as
\[
\rho(t)=\begin{pmatrix}a(t)&c(t)\\c^*(t)&1-a(t)\end{pmatrix},\qquad
c(t)=\tfrac12 e^{-i\omega_0 t}\,e^{-\tfrac12 G(0,t)}.
\]
Defining $r(t)=\sqrt{(a-\tfrac12)^2+|c|^2}$ and using 
$\dot c(t)=(-i\omega_0-\tfrac12\lambda)c(t)$ gives
\[
\frac{dS}{dt}
= -\frac{(a-\tfrac12)\dot a+\mathrm{Re}[c^*\dot c]}{r}
\ln\!\left(\frac{\tfrac12+r}{\tfrac12-r}\right),
\qquad
\mathrm{Re}[c^*\dot c]=-\tfrac14\lambda e^{-G(0,t)},
\quad
\dot a=\Gamma-\lambda a.
\]
Thus the exact entropy-production rate is
\begin{equation}
\label{eq:sigma_exact}
\sigma(t)
= -\frac{(a-\tfrac12)(\Gamma-\lambda a)-\tfrac14\lambda e^{-G(0,t)}}{r(t)}
\ln\!\left(\frac{\tfrac12+r(t)}{\tfrac12-r(t)}\right)
-\beta\,\omega_0\big(\Gamma(t)-\lambda(t)a(t)\big).
\end{equation}

\textbf{Approximate relation in the weak-coherence regime.}In many realistic cases the system, after some damping, is near the maximally mixed point $a\approx \frac{1}{2}$ and coherences are small. Make the small-r expansion $(r<<1)$. Then one may use the quadratic expansion of von Neumann entropy around maximally mixed. For $r\ll 1$, the entropy expansion $S\simeq\ln 2-2r^2 + O(r^4)$ yields
$\frac{dS}{dt}\simeq\frac{\lambda}{2} e^{-G(0,t)}$. Expressing $e^{-G}$ via 
$\mathrm{viol}$ gives the approximate scaling
\begin{equation}
\label{eq:approx_sigma}
\sigma(t_2)\;\approx\;
2\,\lambda(t_2)\,
\frac{\mathrm{viol}(t_1,t_2)^2}
{\sin^2(\omega_0 t_1)\,\sin^2(\omega_0(t_2-t_1))}
\;-\;\beta\,\omega_0\,\dot a(t_2).
\end{equation}
In the small-coherence / near-maximally mixed regime the relation (Approx) holds: $\sigma$ is the sum of a term proportional to $\frac{\lambda(t)\text{viol}^2}{S^2}$ (coherence-driven) and a population/heat term $-\beta \omega_0 \dot{a}$. If population changes are small, $\sigma$ scales approximately like $\text{viol}^2$.
\\
Both KC violation and entropy production stem from the same dynamical origin: the integrated rate $G(0,t)$ governing coherence decay. However, while $\mathrm{viol}$ depends only on $\lambda(t)=\Gamma+\widetilde{\gamma}$, the entropy production rate is additionally sensitive to the individual population contributions $\Gamma$ and $\widetilde{\gamma}$ through $\dot a(t)$. Therefore, a closed functional dependence $\sigma=\sigma(\mathrm{viol})$ does not exist in general. Nevertheless, in regimes where $\dot a\simeq 0$ and $r\ll 1$, Eq.~\eqref{eq:approx_sigma} shows that entropy production is dominated by a term \emph{quadratic} in the KC violation, establishing a direct quantitative link between memory effects and dissipation.


\section{Kolmogorov Consistency Violation and Its Relation to the Leggett--Garg Inequality}
\label{LGI}
We consider a qubit initially prepared in the pure state
\[
\rho(0)=|+\rangle\langle+|,
\qquad
|+\rangle=\tfrac{1}{\sqrt{2}}(|0\rangle+|1\rangle),
\]
evolving under the system Hamiltonian
\[
H=\tfrac{\omega_0}{2}\,\sigma_z,
\]
and interacting with a dissipative environment characterized by time-dependent excitation and relaxation rates
$\widetilde{\gamma}(t)$ and $\Gamma(t)$. The reduced dynamics is assumed to be time-local and is represented by a completely positive dynamical map
$\Lambda(t_b,t_a)$.
As the dichotomic observable we choose
\[
Q=\sigma_x, \qquad Q|q\rangle = q|q\rangle,\quad q=\pm1.
\]

In the computational basis $\{|0\rangle,|1\rangle\}$, the off-diagonal density-matrix element
$c(t)=\rho_{01}(t)$ obeys the equation of motion
\[
\dot c(t)
=
-\Bigl(i\omega_0+\tfrac12[\Gamma(t)+\widetilde{\gamma}(t)]\Bigr)c(t),
\]
whose formal solution for $t_b\ge t_a$ reads
\[
c(t_b)
=
c(t_a)\,
\exp\!\left[-i\omega_0(t_b-t_a)-\tfrac12 G(t_a,t_b)\right],
\quad
G(t_a,t_b)=\int_{t_a}^{t_b}[\Gamma(s)+\widetilde{\gamma}(s)]\,ds.
\]

\medskip

We now derive the relevant two-time correlation functions entering both Kolmogorov consistency and the Leggett--Garg inequality.

\medskip

\textit{Correlation function \(C(0,t_1)\).}  
The two-time correlator between times $0$ and $t_1$ is defined as
\[
C(0,t_1)=\langle Q(0)Q(t_1)\rangle.
\]
In general open-system dynamics, this quantity may be defined either through a Heisenberg-picture (noninvasive) correlator or through a sequential projective measurement protocol. In the present case, these two definitions coincide because the initial state $|+\rangle$ is an eigenstate of $Q$ with eigenvalue $+1$. Consequently, a projective measurement of $Q$ at $t=0$ leaves the state unchanged.

The correlator therefore reduces to a single-time expectation value,
\[
C(0,t_1)
=
\operatorname{Tr}[Q\,\rho(t_1)].
\]
Since $Q=\sigma_x$, this expectation value depends solely on the coherence,
\[
C(0,t_1)
=
2\,\mathrm{Re}\,\rho_{01}(t_1).
\]
Using the solution of the master equation with $c(0)=\tfrac12$, we obtain
\begin{equation}
\label{eq:C01}
C(0,t_1)
=
e^{-\tfrac12 G(0,t_1)}\cos(\omega_0 t_1).
\end{equation}

\medskip

\textit{Sequential correlation function \(C(t_1,t_2)\).}  
We now consider a projective measurement of $Q$ performed at time $t_1$, followed by a second measurement at $t_2>t_1$. The corresponding correlator is defined as
\[
C(t_1,t_2)
=
\sum_{q_1,q_2=\pm1}
q_1 q_2\,
p(q_1\text{ at }t_1,\; q_2\text{ at }t_2).
\]

A measurement at $t_1$ yielding outcome $q_1$ prepares the post-measurement state
\[
\rho(t_1^+)=P_{q_1}
=
\tfrac12
\begin{pmatrix}
1 & q_1\\
q_1 & 1
\end{pmatrix},
\]
which has coherence $c(t_1^+)=q_1/2$.
Evolving this state from $t_1$ to $t_2$ yields
\[
c(t_2)
=
\tfrac{q_1}{2}\,
\exp\!\left[-i\omega_0(t_2-t_1)-\tfrac12 G(t_1,t_2)\right].
\]
The conditional expectation value of $Q$ at time $t_2$ is therefore
\[
\langle Q(t_2)\rangle_{q_1}
=
2\,\mathrm{Re}\,c(t_2)
=
q_1\,e^{-\tfrac12 G(t_1,t_2)}
\cos\!\bigl(\omega_0(t_2-t_1)\bigr).
\]

Substituting this expression into the definition of $C(t_1,t_2)$ and using $\sum_{q_1}p(q_1)=1$ and $q_1^2=1$, we obtain
\begin{equation}
\label{eq:C12}
C(t_1,t_2)
=
e^{-\tfrac12 G(t_1,t_2)}
\cos\!\bigl(\omega_0(t_2-t_1)\bigr).
\end{equation}

\medskip

\textit{Kolmogorov consistency violation.}  
Kolmogorov consistency requires that the marginal probability at time $t_2$ equals the sum of joint probabilities over intermediate outcomes at $t_1$. For dichotomic observables this condition reduces to the multiplicative constraint
$
C(0,t_2)=C(0,t_1)\,C(t_1,t_2).
$
We therefore define the degree of Kolmogorov consistency violation as
\begin{equation}
\label{eq:KC_violation}
\mathrm{viol}_{+}
=
\tfrac12\,
\big|\,C(0,t_1)C(t_1,t_2)-C(0,t_2)\,\big|.
\end{equation}
Substituting Eqs.~\eqref{eq:C01} and \eqref{eq:C12}, we obtain the closed-form expression
\[
\mathrm{viol}_{+}
=
\tfrac12\,
e^{-\tfrac12 G(0,t_2)}
\big|\sin(\omega_0 t_1)\,
\sin\!\bigl(\omega_0(t_2-t_1)\bigr)\big|.
\]

\medskip

\textit{Connection to the Leggett--Garg inequality.}  
A standard three-time Leggett--Garg combination is
\[
K_3
=
C(0,t_1)+C(t_1,t_2)-C(0,t_2),
\]
which satisfies $K_3\le1$ for any macrorealistic theory obeying noninvasive measurability.
Using Eq.~\eqref{eq:KC_violation}, this quantity can be rewritten as
\begin{align}
K_3
&= C(0,t_1)+C(t_1,t_2)-C(0,t_2) \nonumber\\
&= \Bigl[C(0,t_1)+C(t_1,t_2)-C(0,t_1)C(t_1,t_2)\Bigr]
   - \Bigl[C(0,t_2)-C(0,t_1)C(t_1,t_2)\Bigr] \nonumber\\
&=: \mathcal{F}(t_1,t_2)
   + \Bigl[C(0,t_1)C(t_1,t_2)-C(0,t_2)\Bigr] \nonumber\\
&= \mathcal{F}(t_1,t_2)\;\pm\;2\,\mathrm{viol}_{+},
\end{align}
Here the sign is determined by
$\operatorname{sgn}[C(0,t_1)C(t_1,t_2)-C(0,t_2)]$.
\end{widetext}

\end{document}